\pdfoutput=1
\documentclass[aps,preprint,superscriptaddress,onecolumn,longbibliography]{revtex4-1}

\usepackage{graphicx}
\usepackage{epstopdf, epsfig}
\usepackage{subfigmat}
\usepackage{rotating}

\usepackage{tabularx,colortbl}

\newcommand{\PLH}{{\mkern-2mu\times\mkern-2mu}}

\begin{document}

\title{Effect of layout on asymptotic boundary layer regime in deep wind farms}

\author{Juliaan Bossuyt}
\affiliation{Department of Mechanical Engineering, KU Leuven}
\affiliation{Department of Mechanical Engineering, Johns Hopkins University, Baltimore, MD 21218, USA}

\author{Charles Meneveau}
\affiliation{Department of Mechanical Engineering, Johns Hopkins University, Baltimore, MD 21218, USA}

\author{Johan Meyers}
\affiliation{Department of Mechanical Engineering, KU Leuven}

\date{\today}

\begin{abstract}
The power output of wind farms depends strongly on spatial turbine arrangement, and the resulting turbulent interactions with the atmospheric boundary layer. Wind farm layout optimization to maximize power output has matured for small clusters of turbines, with the help of analytical wake models. On the other hand, for large farms approaching a fully-developed regime in which the integral power extraction by turbines is balanced through downwards transport of mean kinetic energy, the influence of turbine layout is much less understood. The main goal of this work is to study the effect of turbine layout on the power output for large wind farms approaching a fully-developed regime. For this purpose we employ an experimental setup of a scaled wind farm with one-hundred porous disk models, of which sixty are instrumented with strain gages. Our experiments cover a parametric space of fifty-six different layouts for which the turbine-area-density is constant, focusing on different turbine arrangements including non-uniform spacings. The strain-gage measurements are used to deduce surrogate power and unsteady loading on turbines for each layout. Our results indicate that the power asymptote at the end of the wind farm depends on the layout in different ways. Firstly, for layouts with a relatively uniform spacing we find that the power asymptote in the fully developed regime reaches approximately the same value, similarly to the prediction of available analytical models. Secondly, we show that the power asymptote in the fully-developed regime can be lowered by inefficient turbine placement, for instance when a large number of the turbines are located in the near wake of upstream turbines. Thirdly, our experiments indicate that an uneven spacing between turbines can improve the overall power output for both the developing and fully-developed part of large wind farms. Specifically, we find a higher power asymptote for a turbine layout with a significant streamwise uneven spacing (i.e. a large streamwise spacing between pairs of closely spaced rows that are slightly staggered). Our results thereby indicate that such a layout may promote beneficial flow interactions in the fully-developed regime for conditions with a strongly prevailing wind direction.
\end{abstract}


\maketitle

\section{Introduction}
\label{s:intro}

Wind turbines are clustered in farms to provide the largest possible cumulative power, within available surface and cost. Inevitably, when turbines are closely spaced together, the momentum deficit in wakes from upstream turbines reduces the available power for downstream ones, while increased turbulence levels result in higher unsteady loading of turbine components. Depending on turbine location, operational control, and inflow conditions, wake induced power losses can be as high as 50\%, compared to a lone standing turbine \cite{nygaard2014wakes}. An important aspect for wind farm design is therefore to better understand the relation between turbine layout, and the resulting wake losses and structural loading.

Analytical wake models that describe downstream advection and expansion of turbine wakes \cite{lissaman1979energy,katic1986simple,bastankhah2014new}, have been useful tools to study the effects of layout on wake losses, e.g. see Ref. \cite{herbert2014review} for a comprehensive overview of optimization studies and Refs. \cite{feng2015solving,parada2017wind,beskirli2017new,abdelsalam2018optimization} for several more recent examples. A classic result is the higher power output for layouts with a larger spacing between streamwise aligned turbines, e.g. a staggered layout as compared to an aligned configuration.

Turbulence resolving numerical simulations, such as LES can be used to study in detail the complex interaction between large wind farms and a turbulent boundary layer \cite{calaf2010large}. Unfortunately, the high computational cost has limited the use of LES for parametric or layout optimization studies. Ref. \cite{bokharaie2016wind} therefore developed a hybrid Jensen-LES optimization procedure, in which the wake coefficients of the Jensen wake model are frequently updated with an LES simulation.  Archer et al. \cite{archer2013quantifying} studied the power output of six different layouts in a LES with a finite size wind farm. In their study, the staggered layout was found to produce the highest power output, showing good agreement with wind tunnel experiments of aligned and staggered wind farms \cite{Chamorro2011Turbulent,Chamorro2011turbulentb,bossuyt2017measurement}. Stevens et al. \cite{stevens2014large} investigated the effect of changing the alignment angle with the wind direction of originally streamwise oriented turbine columns. It was found that an alignment angle smaller than fully staggered can result in an overall higher power output, indicating that a staggered layout is not necessarily the most optimal. While the layout clearly influences the power of the first few rows of the farm, LES results \cite{stevens2014large,stevens2016effects,wu2017flow} and wind tunnel measurements \cite{Chamorro2011turbulentb,bossuyt2017measurement} show that after approximately ten rows, the average row-power becomes independent of row number, thus indicating the approach of a fully-developed regime.

As wind farms become larger and accommodate more rows of turbines, wakes start to encompass the entire farm region. Wake recovery becomes then increasingly more dependent on vertical transport of mean kinetic energy from the high momentum flow above the turbines \cite{calaf2010large,cal2010,verhulst2014large,markfort2018analytical}.
For very large farms, the fully-developed regime can be defined as when the flow becomes statistically independent of downstream turbine row number, and power extraction by turbines becomes fully balanced by overall vertical flux of mean kinetic energy. Under this condition, mean row-power does not change anymore from one row to the next. The vertical transport of mean kinetic energy is governed by Reynolds and dispersive stresses in the shear layer at the top height of the turbines \cite{calaf2010large}, and makes the relation between power output and turbine layout of large farms increasingly more complex.

The power output in the fully-developed regime is traditionally modeled with a top-down description of the horizontally averaged flow field and the vertical interaction between the boundary layer and the horizontally average turbine thrust force applied at hub height \cite{fra92,fra06,calaf2010large,meneveau2012top}. Similarly, Markfort et al. \cite{markfort2018analytical} make the analogy with sparsely-obstructed shear flows, and models the vertical Reynolds shear stress with a Prandtl mixing-length approach. However, due to the horizontally averaged approach, these models cannot take into account the specific effects of turbine layout patterns. As a result, they lead to a single asymptotic value for the mean row-power output of an infinite wind farm, solely as a function of turbine-area-density. However, so far, it is not clear how for a fixed turbine density, the turbine layout influences the power-asymptote in the fully-developed regime. Moreover, it is unclear if the asymptote from the top-down approach should be considered as an upper limit, or if higher efficiencies are possible with, for instance, arrayed layouts with constant spacing.

Similarly, for reasonably small spanwise spacings (e.g. inter-turbine distance smaller than $6D$, where $D$ is the turbine rotor diameter), LES studies \cite{yang2012computational,stevens2016effects} and experiments \cite{bossuyt2016measuring} for aligned and staggered array configurations found that in the fully-developed limit the mean power was almost independent of the actual turbine arrangement. Specifically, the staggered layout was found to result in nearly the same power output as the algined layout with the same turbine-area-density $S_x\PLH S_y$, despite the difference in streamwise spacing $S_x$ (with $S_y$ the spanwise spacing). 

Nevertheless, periodic LES studies of infinite farms \cite{verhulst2014large,chatterjee2018contribution} have indicated that the spacing between turbines in an aligned or staggered layout does influence the turbulent structures responsible for vertical transport of mean kinetic energy, and that these scales can be an order of magnitude larger than the turbine diameter. Chatterjee and Peet \cite{chatterjee2018contribution} found specifically that by increasing the turbine spacing, one can increase the turbulent length scales responsible for downwards transport, and therefore potentially benefit the overall wind farm efficiency. The question thus arises, how the power output in the fully-developed regime can be increased by selecting turbine arrangements that optimally stimulate the structure of turbulent scales responsible for energy transfer to turbines.

The idea that local flow interactions between closely spaced drag objects can increase the integral drag force of the group of roughness elements (which in the context of wind farms can be considered directly related to power output) has been observed before in the literature \cite{taddei2016characterisation}, and highlights an interesting concept that may help to improve overall wind farm efficiency. For instance, McTavish et al. \cite{mctavish2014experimental} showed the potential of this concept with a wind tunnel experiment of three scaled turbines closely placed together, e.g. placing one turbine just downstream and in the middle between two others. This concept aims at increasing the overall power by benefiting of the local flow acceleration between the two upstream turbines, which is an effect that would not be readily captured by conventional wake models.

\begin{table}
	\centering
	\begin{tabular}{lllll}
		Authors&\# turbines&$D$ [m]&$Re_D$&Layouts \\ \hline 
		Cal et al. \cite{cal2010}&9 WT&0.12&$6.4\PLH10^4$&AL: $3\PLH3$\\
		Corten et al. \cite{Corten2004Turbine}&28 WT&0.25&$7.5\PLH10^4$&AL: $8\PLH3$, $7\PLH4$, $14\PLH2$, ST: $9\PLH3$\\
		Chamorro and Port\'e-Agel \cite{Chamorro2011Turbulent}&30 WT&0.15&$2.0\PLH10^4$&AL: $10\PLH3$\\
		Chamorro et al. \cite{Chamorro2011turbulentb}&30 WT&0.128&$2.2\PLH10^4$&ST: $10\PLH3$\\
		Markfort et al. \cite{markfort2012turbulent}&36 WT&0.128&$2.2\PLH10^4$&AL: $12\PLH3$, ST: $12\PLH3$\\
		Charmanski et al. \cite{charmanski2014physical}&91 PD+9 WT&0.25&-&AL: $5\PLH5$, $9\PLH5$, $14\PLH5$, $19\PLH5$\\
		Theunissen et al. \cite{Theunissen2015Experimental}& 80 PD&0.025&$4.1\PLH10^4$&Horns Rev (rhomboid:$10\PLH8$)\\
		&&&&ST: $8\PLH10$, $8\PLH5$, AL: $6\PLH8$\\
		&&&&4 wind directions/case\\
		Bossuyt et al. \cite{bossuyt2017measurement}&100 I-PD&0.03&$2.1\PLH10^4$&AL: $20\PLH5$, ST: $20\PLH5$\\ 
		&&&&+4 intermediate alignments
	\end{tabular}
	\caption{Summary of large wind farm experiments in the literature. The Reynolds number $Re_D$ is estimated based on the diameter and the documented hub-height velocity in the experiment. Layouts are noted by \# spanwise rows $\times$ \# of streamwise columns. The following abbreviations are used: WT = scaled wind turbine model, {PD = scaled porous disk model},  \mbox{I-PD = instrumented porous disk model}, {D = diameter}, AL = aligned layout, ST = staggered layout.}
	\label{t:wfexperiments}
\end{table}

Wind tunnel experiments allow to measure many layouts relatively cheaply in well defined flow conditions, and are thus ideal for parametric studies. However, due to scaling-related challenges, experiments have mostly studied smaller farms, and also only few layouts, such as aligned or staggered. An overview of wind tunnel experiments in the literature is presented in \mbox{table \ref{t:wfexperiments}}.

Our goal is to explore the potential of non-uniform and large streamwise turbine spacings, with the aim to improve overall farm performance in both the entrance and fully-developed part of a large farm. We aim at providing new insights that can inspire and motivate future LES studies, which are currently too expensive for large parametric studies, but are especially valuable to study the detailed turbulent flow interactions. In this paper, we employ an experimental setup of a scaled wind farm with one-hundred porous disk models and twenty spanwise rows, to study farms that approach a fully-developed regime. Making use of the experimental capability to measure many layouts at a relatively low cost, we perform a parametric study of fifty-six different turbine layouts, of which an example is shown in figure \ref{f:photo2}. The experimental setup was previously designed and validated by Bossuyt et al. \cite{bossuyt2017measurement}. Thanks to the instrumentation of sixty porous disk models with strain gages, the measurements contain detailed information about the mean surrogate power in each row and the temporal statistics, related to unsteady loading. All experiments are performed for the same inflow conditions and one fixed wind direction, to provide a well defined setup and enable clear comparisons between layouts. In first instance, these results can thus be applicable to wind farms with a dominant wind direction.

Section \ref{s:expsetup} of this paper describes the experimental setup, and provides a validation of the porous disk instrumentation by comparing with hot-wire measurements for the two most well documented layouts in the literature: an aligned and staggered configuration. In Section \ref{s:wf_measurements}, the measurement results for all fifty-six layouts are presented and discussed. Finally, in section \ref{s:wf_discussion}, the overall farm performance of each layout is compared and discussed.

\section{Experimental setup}
\label{s:expsetup}

\begin{figure}
	\centering
	\includegraphics[width = 0.7\textwidth]{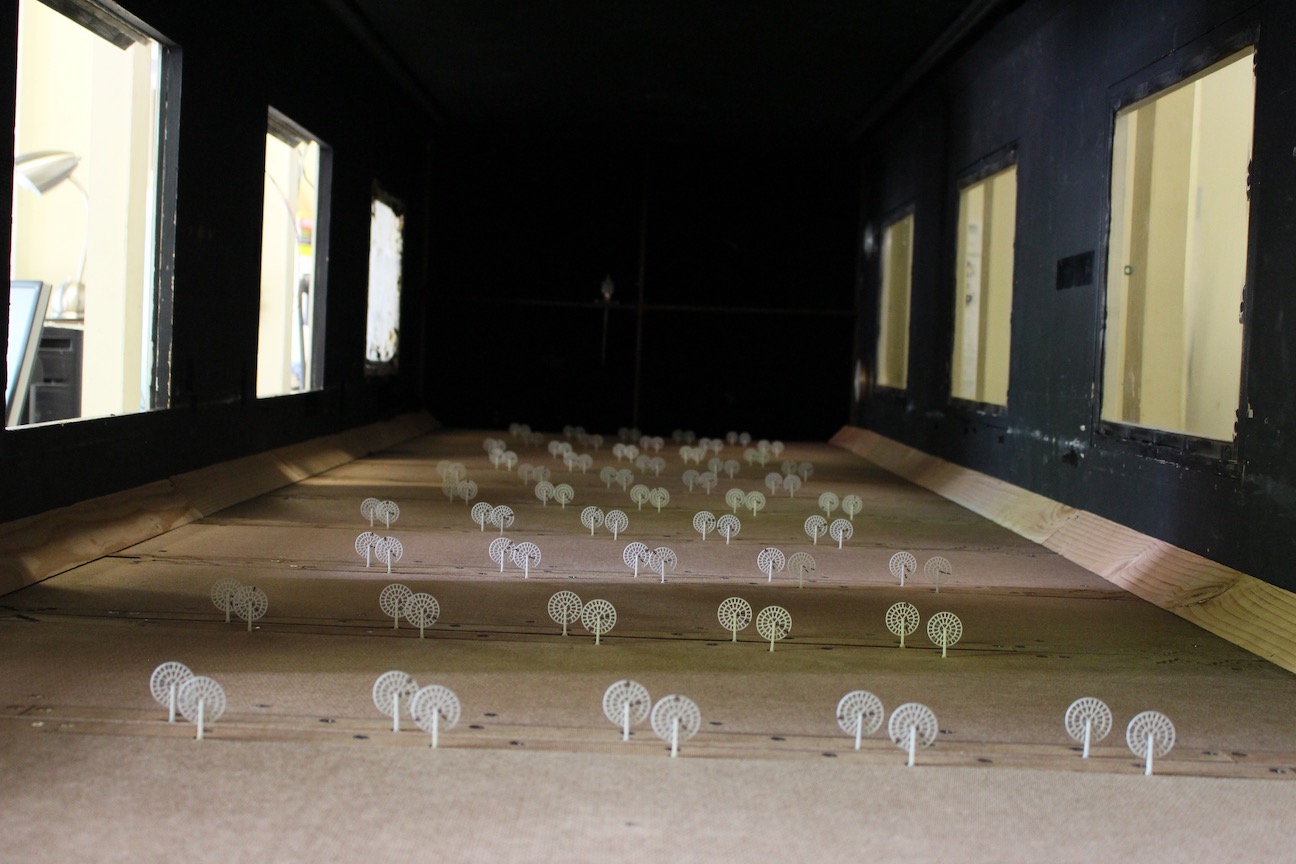}
	\caption{A photograph of the \textit{NU2-C3} layout with a spanwise shift of $1D$ in the wind tunnel.}
	\label{f:photo2}
\end{figure}

Figure \ref{f:photo2} shows a photograph of the experimental setup in the wind tunnel. In section \ref{ss:microwf} and \ref{ss:mwf}, we motivate and describe the original design and validation of the micro wind farm model by Bossuyt et al. \cite{bossuyt2017measurement}. In section \ref{ss:layout} an overview and description of all measured layouts is presented. In section \ref{ss:validation} we provide a validation of the porous disk instrumentation with detailed hot-wire measurements for the aligned and staggered layout.

\subsection{Porous disk modeling}
\label{ss:microwf}

For the purpose of this study, the experimental setup should model a wind farm large enough to approach a fully-developed regime. The required wind farm size depends on the boundary layer conditions, turbine spacing, and how the criteria for a fully-developed condition are defined \cite{wu2017flow}. As a first order approximation, we assume that the mean-row power can be considered an appropriate indicator for approaching a fully-developed condition. Field measurements of the Horns Rev wind farm \cite{barthelmie2011flow} and laboratory experiments of scaled farms \cite{markfort2012turbulent} found that the development to a fully-developed regime, as indicated by the evolution of mean row power in the field measurement or rotor speed in the experiment, required on the order of ten turbine-rows. Based on the scaling argument presented by Markfort et al. \cite{markfort2018analytical}, one can expect that up to twenty rows are necessary for a farm with realistic spacings $S_x=7D$, $S_y=5D$ and thrust coefficient $C_T=0.75$. Previous experiments \cite{bossuyt2017measurement} with the experimental setup used in this study show that the mean surrogate power reaches an approximate plateau for both the aligned and staggered layout around the seventeenth row. To fit a scaled farm with twenty rows in a wind tunnel test-section with a typical length on the order of $5-10\,{\rm m}$, the scaled turbine model must have a diameter as small as $D=0.025 - 0.07{\rm \, m}$. 

The flow over a turbine blade operating in the atmospheric boundary layer is characterized by very large Reynolds numbers, e.g. the chord based Reynolds number can exceed $Re_c \sim 10^7$. Without the use of a pressurized wind tunnel, for instance see the experiments by Miller et al. \cite{miller2016model}, flow similarity is impossible due to scaling limitations by compressibility effects. As a result, scaled wind turbines that operate at a lower Reynolds number cannot reach the performance of full-scale ones. For instance, the turbine efficiency is directly related to the local lift and drag forces over the miniature blades, which become increasingly viscosity dependent for small chord lengths and lower wind speeds. Researchers have therefore designed scaled rotors that perform better at lower Reynolds numbers \cite{Corten2004Turbine,Medici2006Measurements,bastankhah2017new,coudou2018experimental}, but do not follow geometric similarity. As a result, small scale turbines typically operate at a higher blade loading (e.g. use larger blade chords) and at a lower tip speed ratio. Chamorro et al. \cite{chamorro2012reynolds} found that wake properties become especially Reynolds dependent for Reynolds numbers lower than $Re_D < 4.8 \times 10^4$. 

The design challenges of scaling turbines for wind tunnel studies of large farms have motivated the development of static porous disk models \cite{Aubrun2013Wind, charmanski2014physical,Theunissen2015Experimental,bossuyt2017measurement}, in analogy to the numerical approach of actuator disk models in LES \cite{mikkelsen2003actuator,calaf2010large}. Porous disk models are designed to exert the same integral thrust force on the flow, and to create an equivalent turbulent wake by mimicking the flow-through behavior of a wind turbine rotor. Porous disk models are drag based, instead of lift, and the local flow separation points of the flow over the porous grid are fixed by sharp edges. They can thus be expected to be less Reynolds number dependent than scaled rotors, for which performance depends on local lift forces. By providing significant flow through, porous disk models don't exhibit bluff-body vortex shedding (as shown by Ref. \cite{castro1971wake} for a porosity higher than $0.4$), in agreement with a typical wind turbine wake. Therefore, wind tunnel measurements of porous disks in a turbulent boundary layer are considered possible for Reynolds numbers as lows as \mbox{$Re_D = 2 - 3 \times 10^4$ \cite{lim2007bluff}}. 

It is important to note that the wake of a porous disk does not contain wake rotation and other specific blade signatures, such as helical tip vortices. However, in a turbulent boundary layer, these features have been found to be rapidly overwhelmed by ambient turbulence after a downstream distance of several rotor diameters, e.g. the far wake region \cite{Aubrun2013Wind,camp2016mean,bossuyt2017measurement}. A detailed analysis with PIV measurements confirmed that outside the near wake region, the vertical transport of mean kinetic energy is represented fairly well, making porous disk models suitable for studies of large wind farms and their vertical interaction with the boundary layer \cite{camp2016mean}. Theunissen et al. \cite{Theunissen2015Experimental} demonstrated the use of small scale porous disk models with a diameter of $0.025\,{\rm m}$ and a Reynolds number of $Re_D = 4.1 \PLH 10^4$, for a wind tunnel study of the Horns Rev wind farm with $80$ models. 

Conventional scaled turbine models allow to measure electrical power from a generator \cite{Corten2004Turbine}, aerodynamic rotor torque \cite{kan10}, or rotational speed of the blades \cite{Chamorro2011turbulentb}, as a measure for turbine performance. Porous disk models are static, and don't convert the dissipated kinetic energy to useful electrical power. Therefore, an estimate for turbine performance must be obtained in an indirect way. Initially, studies focused on measurements of the velocity field \cite{charmanski2014physical}, or the integral drag force of the entire scaled farm \cite{Theunissen2015Experimental}. More recently, Bossuyt et al. \cite{bossuyt2017measurement} instrumented individual porous disk models with strain gages, to measure the instantaneous integral thrust force. This technique conveniently allows time-varying measurements, which can be used to reconstruct the spatially averaged incoming velocity time signal, and a surrogate power signal for each instrumented model. The temporal resolution has allowed the study of spatio-temporal characteristics of turbine surrogate power signals in large wind farms \cite{bossuyt2017measurement,bossuyt_meneveau_meyers_2017}.

\begin{figure}
	\centering
	\includegraphics[width = \textwidth]{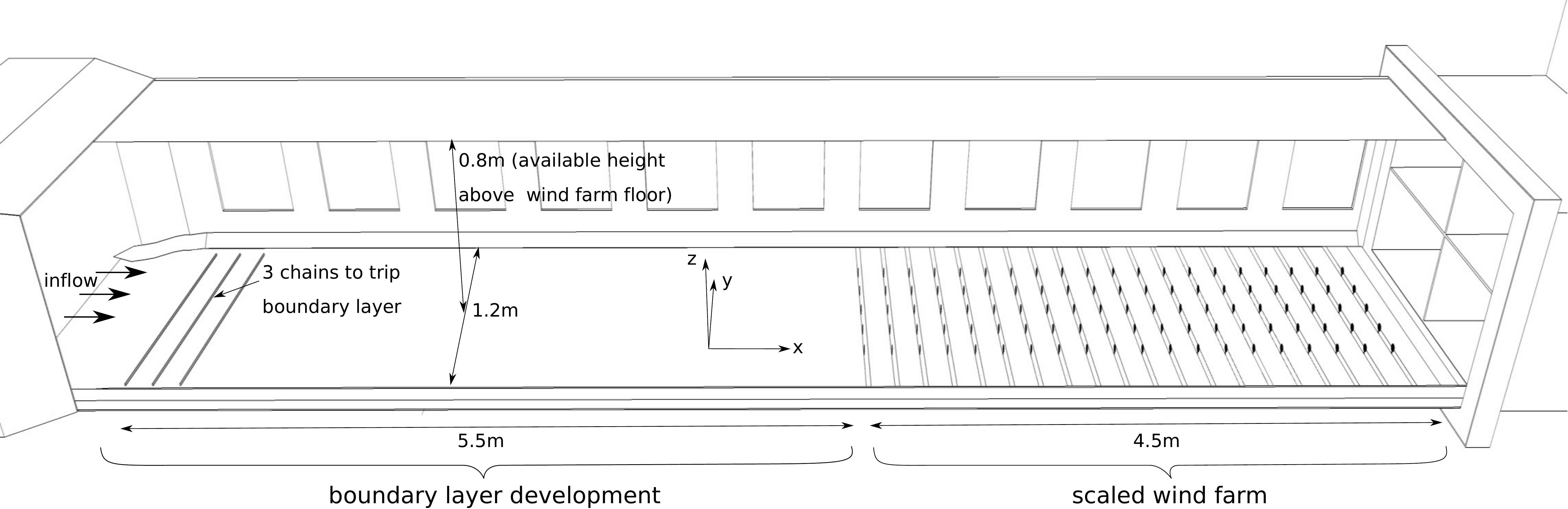}
	\caption{Illustration of the wind tunnel test-section with the scaled wind farm model at the end, and a fetch region to develop the turbulent boundary layer. Image not to scale.}
	\label{f:windtunnel}
\end{figure}

In this study, we employ the instrumented porous disk approach to allow the measurement of the time-dependent thrust forces on sixty porous disk models of a scaled wind farm with one-hundred models in total, and for fifty-six different layouts. The wake of a porous disk is clearly an approximation of a real wind turbine wake. Nevertheless, the thrust force and mean velocity deficit are modeled fairly well, especially in a turbulent flow and outside of the near wake region. Porous disk wakes are well characterized, and the analogy with the numerical actuator disk model enables comparison with LES \cite{bossuyt2018large}. Taking into account these limitations, the porous disk method is considered suitable to model the large scale interaction with the boundary layer, which is induced by the thrust forces, the resulting turbulent wakes and shear in the mean velocity profile. These features are characteristic for shear-obstructed flows and are main mechanisms governing farm performance and downwards transfer of mean kinetic energy in the fully-developed regime \cite{markfort2018analytical}.

\subsection{Micro wind farm}
\label{ss:mwf}

In this section, we present a brief overview of the experimental setup, which was originally designed by Bossuyt et al. \cite{bossuyt2017measurement}. For a detailed description of the design we refer to \mbox{Ref. \cite{bossuyt2017measurement}}. The wind tunnel experiments are performed in the Corrsin Wind Tunnel at Johns Hopkins University, which has a test-section of $10 \times 1.2 \times 0.9 \,{\rm m}$, following a primary contraction of $25:1$, and a secondary of $1.27:1$, to generate a clean inflow with a measured turbulence level of $TI_u \approx 0.12\%$. The test-section width increases downstream to compensate for boundary layer development along the walls. The experiments make no use of any turbulence grids, and instead let the clean inflow develop a turbulent boundary layer over the wind tunnel floor in the first half of the test-section, after being tripped at the entrance by three chains attached to the bottom surface. The experimental setup is described in the schematic shown in figure \ref{f:windtunnel}.

\begin{figure}
	\centering
	\includegraphics[width = 0.45\textwidth]{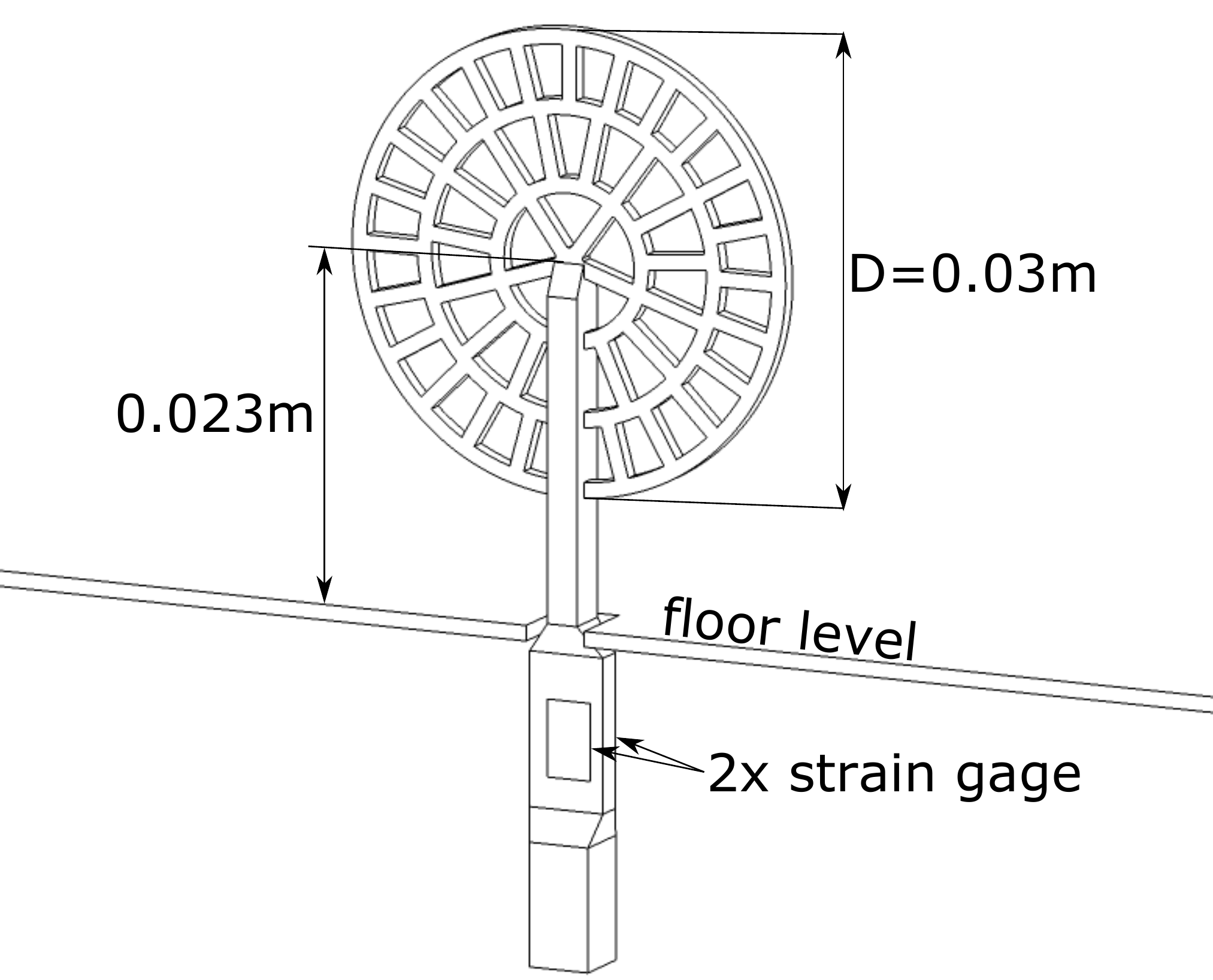}
	\caption{Porous disk model}
	\label{f:porousdisk}
\end{figure}

To fit a scaled wind farm with twenty rows in the wind tunnel test-section, a rotor diameter of  $D=0.03{\rm \, m}$ was selected. Compared to a full-scale wind turbine with a diameter of $100{\rm \,m}$, the porous disk model has geometric scaling ratio of $1:3333$. The porous disk design by Bossuyt et al. \cite{bossuyt2017measurement} is shown in figure \ref{f:porousdisk}.  The porosity of the disk was selected to match a realistic trust coefficient, which was measured to be $C_T = 0.75 \pm 0.04$. Hot-wire measurements in the wake have shown that the normalized mean velocity profile at a downstream distance of $3D$ is in good agreement with results in the literature for scaled turbine models \cite{bossuyt2017measurement}. The bending moment of the model tower, which is a direct result of the integral thrust force on the disk, is measured by two SGD-3/350-LY11 strain gages in a half-bridge configuration to improve accuracy. The time-dependent thrust force is reconstructed from the strain signals by modeling the structural response as a harmonic oscillator for the first and dominant natural frequency of the model. The structural model requires three parameters for each instrumented porous disk. The spring constant is calibrated for all models individually by making use of an automated calibration unit. The damping coefficient was measured from the impulse response, a single value of $\zeta = 0.03$ is used for all models. The natural frequency is on average $\overline{f_n} \approx 200\,{\rm Hz}$, and is determined for each model from the peak in the strain signal spectrum.

With the known thrust coefficient $C_T$, it is possible to estimate the spatially averaged incoming velocity $\langle U \rangle(t)$ from the force measurements $F(t)$, by making use of \mbox{$F(t) = \rho \langle U \rangle ^2 (t) C_T A/2$}, with $\rho$ the density of air and $A = \pi D^2 / 4$ the rotor swept area. The reconstructed velocity can be considered as a uniform incoming velocity which would provide the same measured thrust force. From the reconstructed velocity and by considering a realistic power coefficient $C_P$, one can estimate a representative power signal \mbox{$P(t) = \rho \langle U \rangle ^3 (t) C_P A / 2$}, here refered to as surrogate power output.  In this study, we compare surrogate power values normalized by the surrogate power of the first row, such that results are independent of the specific power coefficient. This methodology can be used for wind turbines operating in the below-rated regime, for which performance is maximized, and the resulting thrust and power coefficient is nearly constant. The frequency response of the thrust force measurements was determined by comparing the spectrum of reconstructed signals with that of a simultaneously measured hot-wire signal. The frequency response was observed to reach up to the natural frequency of the model, and captures the spatial filtering of the turbulent velocity field by the porous disk in the experiments. The turbulence intensity of the reconstructed velocity signal is thus directly representative for the unsteady loading of a porous disk model. It is important to note that reconstruced velocities are spatially filtered, e.g. they follow from an integral over the disk area, such that variances will differ from the unfiltered quantities. The fluctuations of reconstructed velocity or power thus contain turbulent scales similar and larger than the disk diameter.

The micro wind farm consists of one-hundred instrumented porous disk models, organized in twenty rows and five columns. The signal from the sixty porous disk models in the central three columns was measured, to use the instrumentation resources on those models least affected by wind farm border effects. The strain gage signals were measured by Omega iNET-423 voltage input cards with i512 wiring boxes, and one \mbox{Omega iNET-430} 16bit A/D converter. The internal $4\,{\rm kHz}$ low-pass filters are used to reduce high frequency noise from each strain signal. The large number of simultaneous strain gage measurements limited the sampling frequency per model to $866\, {\rm Hz}$, which is lower than required by the Nyquist criteria of the low-pass filter. However, measurements for a single model have validated that the aliasing error is relatively small for the frequency range of interest: $0-200\,{\rm Hz}$ (which is limited by the natural frequency of the model and signal to noise ratio). As indicated in figure \ref{f:porousdisk}, the strain gage sensors are located below the wind farm floor, which reaches $0.1{\rm \,m}$ above the test-section floor. The height of the cross section above the wind farm floor is $0.8{\rm \, m}$.

The measurement results for the \textit{U-C1} layout series (all layouts are introduced in the next section, see \mbox{figure \ref{f:schematic_wt}} for an overview) are those documented by Ref. \cite{bossuyt2017measurement}, which used a measurement time between $5$ and $15$ minutes. For all other layouts, new experiments were performed with a measurement time of approximately $7$ minutes. The acquisition time is thus over $3$ to $9 \times 10^4$ the largest integral time scale ($\approx\,9\,{\rm ms }$) measured for the incoming boundary layer, so that very well converged statistics are obtained for all layouts. While the statistical uncertainty is minimized by a significant measurement time, the strain gages introduce an uncertainty for mean quantities due to potential systematic errors. The measurement uncertainties are estimated from a propagation analysis, and are $\delta U_i/U_1 = \pm 0.03$ for reconstructed velocities measured by a porous disk, $\delta P_i/P_1 = \pm 0.08$ for individual surrogate power signals, $\delta P_i/P_1 = \pm 0.05$ for the row averaged surrogate power signals, $\delta P_i/P_1 = \pm 0.01$ for the surrogate power averaged over 19 rows, $\delta P_i/P_1 = \pm 0.02$ for the surrogate power averaged over 4 rows, and $\delta TI/TI = \pm 0.03$ for turbulence intensities as calculated from the reconstructed velocity.

At a location of $0.21\,{\rm m}$, or $7D$, upstream of the wind farm, the boundary layer height was measured to be $\delta_{99}=0.16\,{\rm m}$, and corresponds to four times the porous disk top height. The roughness length, as estimated by extrapolating the measured log-law velocity profile, is $z_0 = 0.9 \times 10^{-2}\,{\rm mm}$. With a geometric scaling ratio of $1:3333$, the corresponding full-scale roughness is $z_0 = 0.03\,{\rm m}$, comparable to a moderately rough boundary layer. The measured friction velocity was $u_\tau = 0.6\,{\rm m/s}$, obtained from the slope of the mean velocity profile and by assuming a von K\'arm\'an constant of $\kappa = 0.4$. Profiles of measured mean velocity, turbulence intensity, integral length scale, and velocity spectra of the incoming boundary layer flow are documented by Ref. \cite{bossuyt2017measurement}. The blockage ratio of the wind farm is small. Without taking the porosity of the disks into account, the ratio of frontal area covered by porous disk models to the area of the cross section in the wind tunnel is $0.4\%$ for an aligned layout and $0.8\%$ for a staggered layout, so that we do not expect significant blockage effects.

\subsection{Layout description}
\label{ss:layout}

\begin{figure}
	\centering
	\includegraphics[width = 0.8\textwidth]{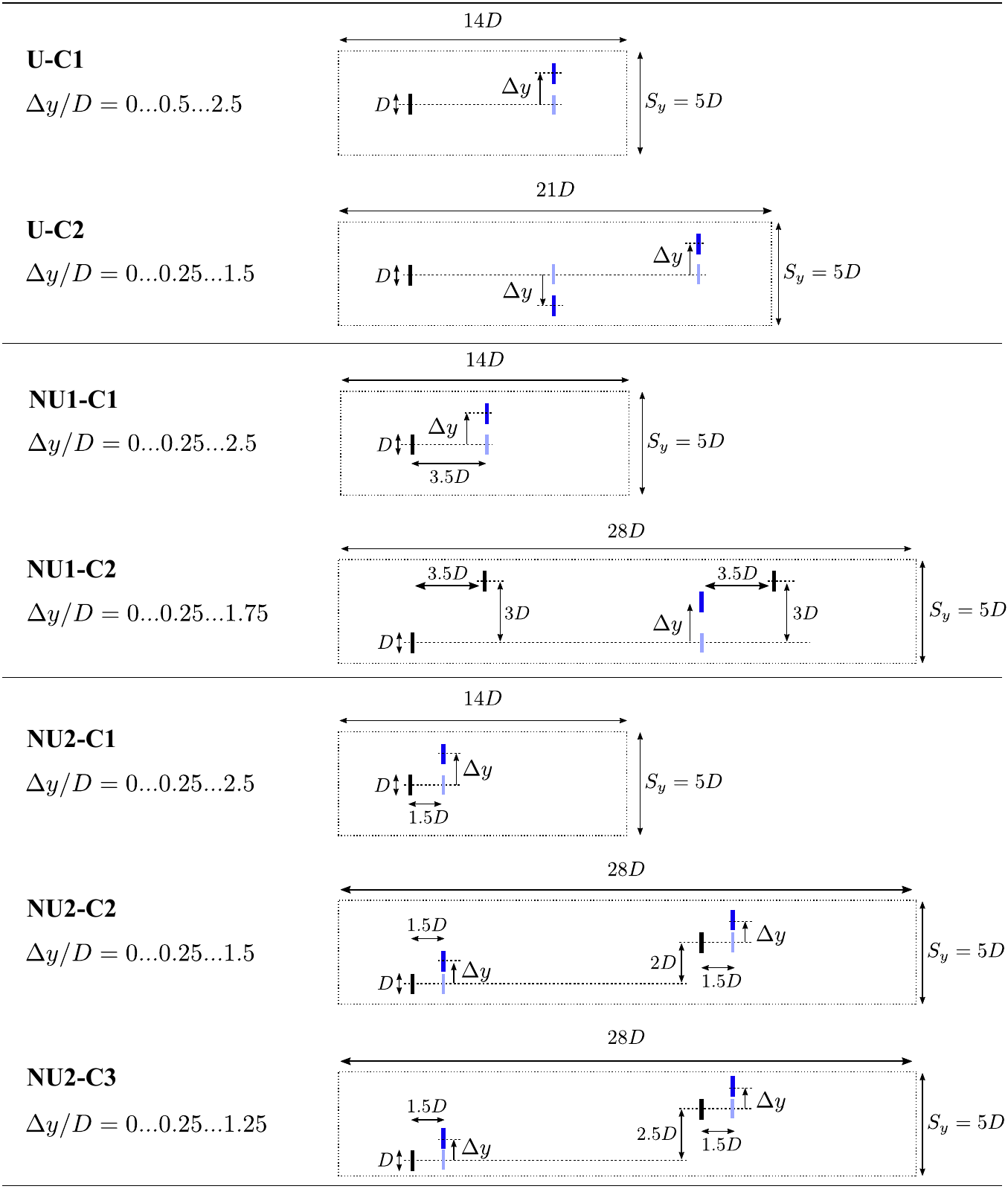}
	\caption{An overview of the studied wind farm layout patterns. Each series consists of a number of layouts, by sliding the indicated (blue) porous disk models in the spanwise direction, over the specified range for $\Delta y$. The total wind farm layout results from repeating the displayed unit cells five times in the spanwise direction, and until a total of twenty rows is reached in the streamwise direction.}
	\label{f:schematic_wt}
\end{figure}

\begin{figure}
	\centering
	\includegraphics[width = \textwidth]{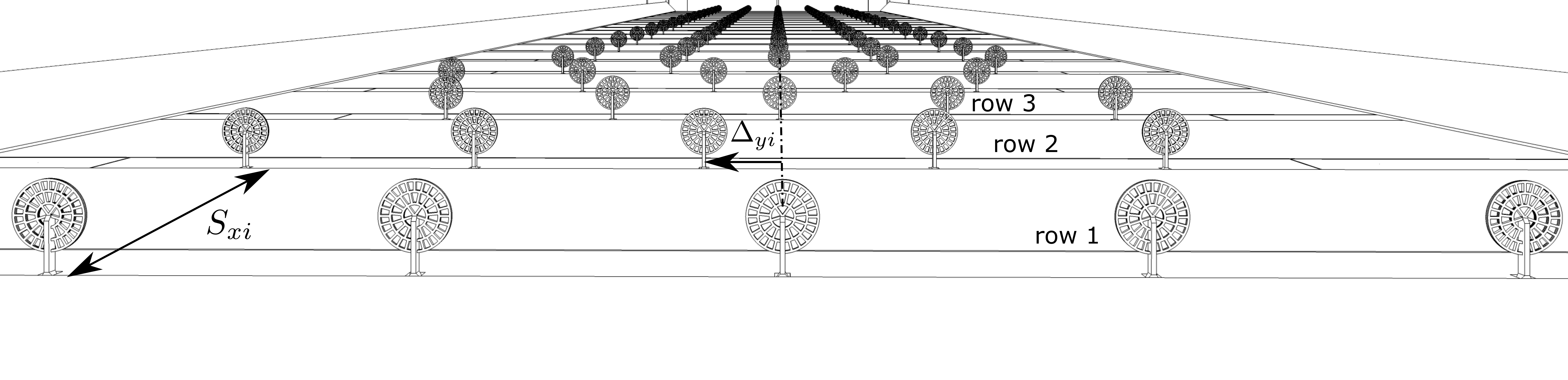}
	\caption{The layout of the wind farm consists of twenty rows with five porous disk models each, and can be altered by sliding rows in the spanwise direction (i.e. $\Delta {yi}$ for each row $i$), or by changing the streamwise spacing between rows (i.e. $S_{xi}$ between row $i$ and $i+1$). Data were acquired from the three instrumented porous disk models in the middle of each row.}
	\label{f:layout}
\end{figure}

The fifty-six wind farm layouts studied in this work are presented in figure \ref{f:schematic_wt}. The total wind farm layout results from repeating the displayed layout-unit cells over the entire wind farm. Specifically, five times in the spanwise direction, and five to ten times in the streamwise direction, depending on the number of rows in the unit cell (e.g. two, three or four). For each layout the same area is occupied in the wind tunnel, so that the area-density of porous disk models is constant, e.g. an area of $7D \times 5D = 35D^2$ for each porous disk model. As indicated in figure \ref{f:layout}, the wind farm arrangements are configured by changing the intermediate streamwise spacing $S_{xi}$, and by sliding rows in the spanwise direction with $\Delta_{yi}$. It is noted that the spanwise spacing between models in each row is always $S_y/D = 5$. A layout with a zero spanwise shift, is referred to as 'aligned', and a layout with a maximal spanwise shift is referred to as 'staggered'.

The first series of layouts considers a uniform streamwise and spanwise spacing. For this layout series, two cases are considered. The first case, \textit{U-C1}, consists of six layouts (originally measured by Bossuyt et al. \cite{bossuyt2017measurement}), which range from aligned to staggered, by sliding the even rows in steps of $0.5D$. The second case, \textit{U-C2}, considers double staggering, for which each third row is slid in the other direction than each second row. 

The second layout series consists of an uneven streamwise spacing which alternates between $S_x/D = 3.5$ and $S_x/D =10.5$. Again two cases are considered. The first case, \textit{NU1-C1}, follows the original approach of varying an aligned layout to a staggered configuration, by sliding the even rows. The second case, \textit{NU1-C2}, moves every third row in a pattern of four rows, while the second and fourth rows have a fixed spanwise shift of $3D$ compared to the first row in the pattern.

The third layout series considers a more extreme non-uniform streamwise spacing, which alternates between $S_x/D = 1.5$ and $S_x/D = 12.5$. Three cases are considered, for which the first, \textit{NU2-C1}, follows again the original aligned to staggered approach. The second case, \textit{NU2-C2}, repeats a pattern of four rows, for which the last two rows together are staggered with a distance of $2D$ compared to the first two. The even rows are moved in steps of $0.25D$. The third layout case, \textit{NU2-C3}, follows a similar approach, but now the first and third row are spaced $2.5D$ in the spanwise direction.

\subsection{Validation}
\label{ss:validation}

\begin{figure}
	\centering	
	\begin{subfigmatrix}{4}
		\centering	
		\subfigure[]{\includegraphics[trim = 0mm 0mm 0mm 2mm ,clip,width = \textwidth, keepaspectratio]{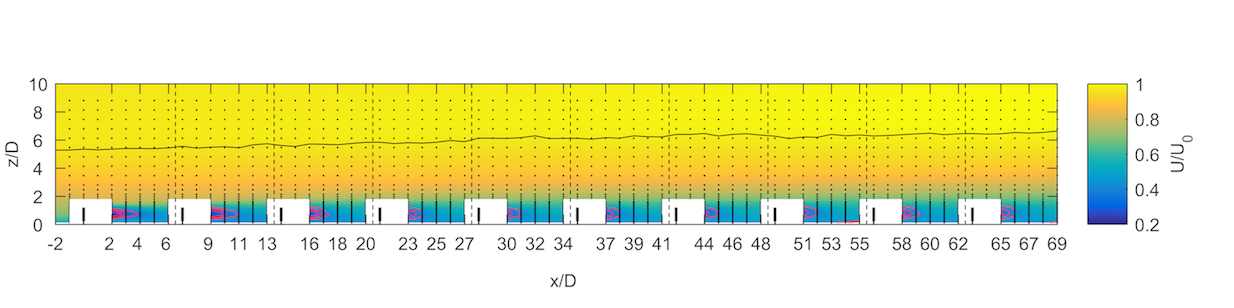}}
		\subfigure[]{\includegraphics[trim =0mm 0mm 0mm 2mm  ,clip,width = \textwidth, keepaspectratio]{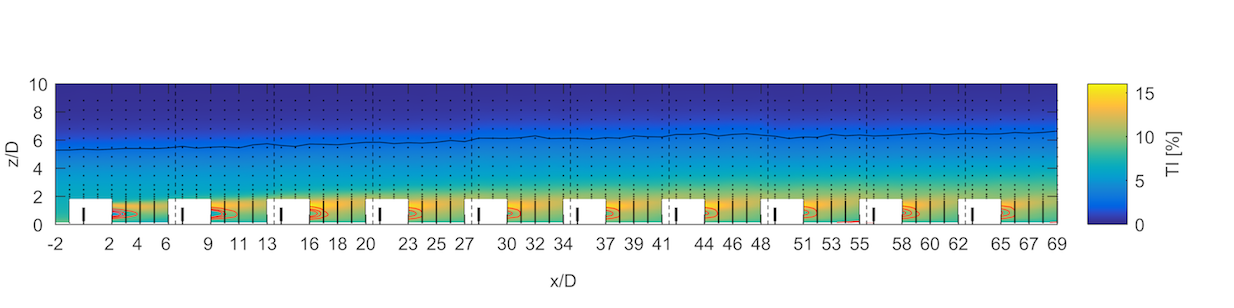}}
		\subfigure[]{\includegraphics[trim = 0mm 0mm 0mm 2mm ,clip,width = \textwidth, keepaspectratio]{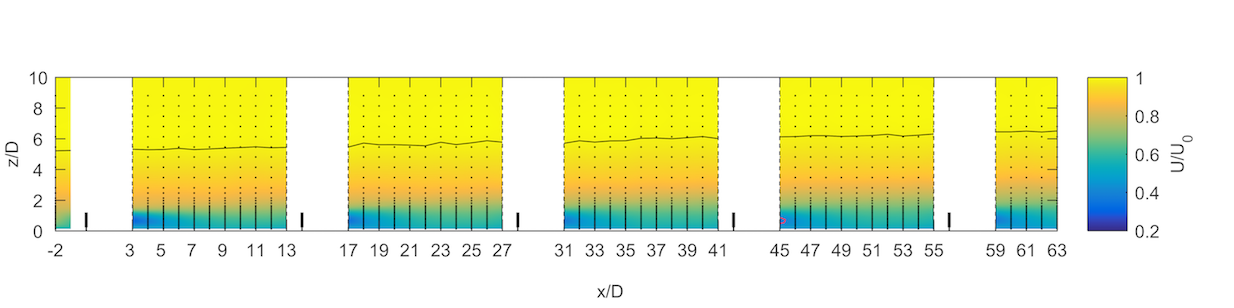}}
		\subfigure[]{\includegraphics[trim =0mm 0mm 0mm 2mm  ,clip,width = \textwidth, keepaspectratio]{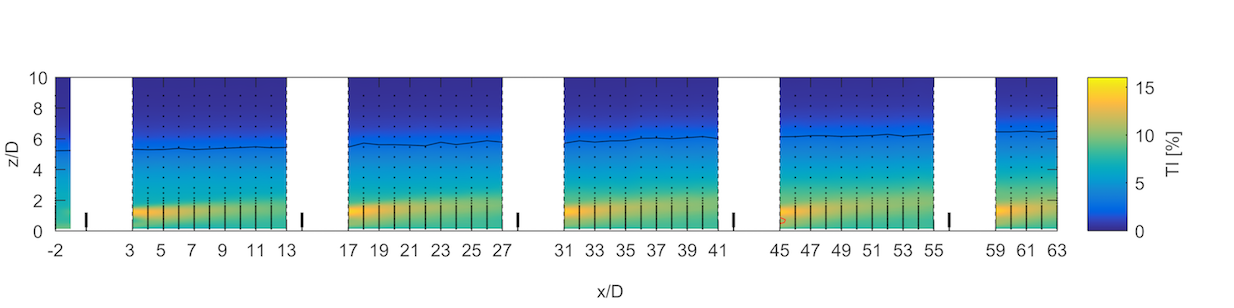}}
	\end{subfigmatrix}
	\caption{Hot-wire measurements of the mean streamwise velocity and turbulence intensity for an aligned layout (a,b), and for a staggered layout (c,d). Vertical dotted lines indicate the separate measurement series, and the solid black line shows where the mean velocity reaches $99\%$ of the freestream velocity, as an indication for the boundary layer height. The hot-wire probe was calibrated to measure velocities as low as $3\,{\rm m/s}$. In the turbulent low-momentum wakes, red contour-lines indicate where the measured velocity values were lower than this threshold for a percentage of $1\%$, $5\%$, or more than $10\%$ of the total measurement points.}
	\label{hw_aligned}
\end{figure}

The micro wind farm setup used in this study was previously successfully used to measure the layout-dependent spatio-temporal characteristics of turbine power outputs \cite{bossuyt2017measurement}, and to validate an LES code, showing good agreement for mean row-power values \cite{bossuyt2018large}. In this section we extend this original validation of the experimental setup with a comparison between mean velocities deducted from the porous disks and direct hot-wire measurements. The horizontal velocity component, in a vertical (X--Z) plane through the central column of the wind farm, was measured for two layouts: aligned and staggered. The measurements were performed with an in-house built one-component hot-wire probe, positioned in each measurement point with an in-house built automated traversing system. The measurements cover the first ten rows. The acquisition was done with a TSI IFA-300 Constant Temperature Anemometer hot-wire system and a PCI-PD2-MFS-8-1M/12 data acquisition card. The velocity at each point was filtered with an analog low pass filter of $5\, {\rm kHz}$ and acquired for $52.4$ seconds at a sampling frequency of $10\,{\rm kHz}$. The hot-wire measurements were acquired over several independent measurement series covering several days, and stitched together based on a reference pitot measurement in the free-stream. These pitot measurements where also used during each measurement to regularly re-calibrate the hot-wire probe \cite{talluru2014calibration}. Contours of the mean velocity and turbulence intensity ($TI_u = \sqrt{\overline{u^{'2}}} / U_0$ are shown in figure \ref{hw_aligned}, with $U_0$ the free-stream velocity, $u^{'}$ the velocity fluctuation and the temporal mean denoted with the overline, such that: $u = \overline{u} + u^{'}$.

\begin{figure}
	\begin{subfigmatrix}{2}
		\subfigure[]{\includegraphics[width = 0.8\textwidth, keepaspectratio]{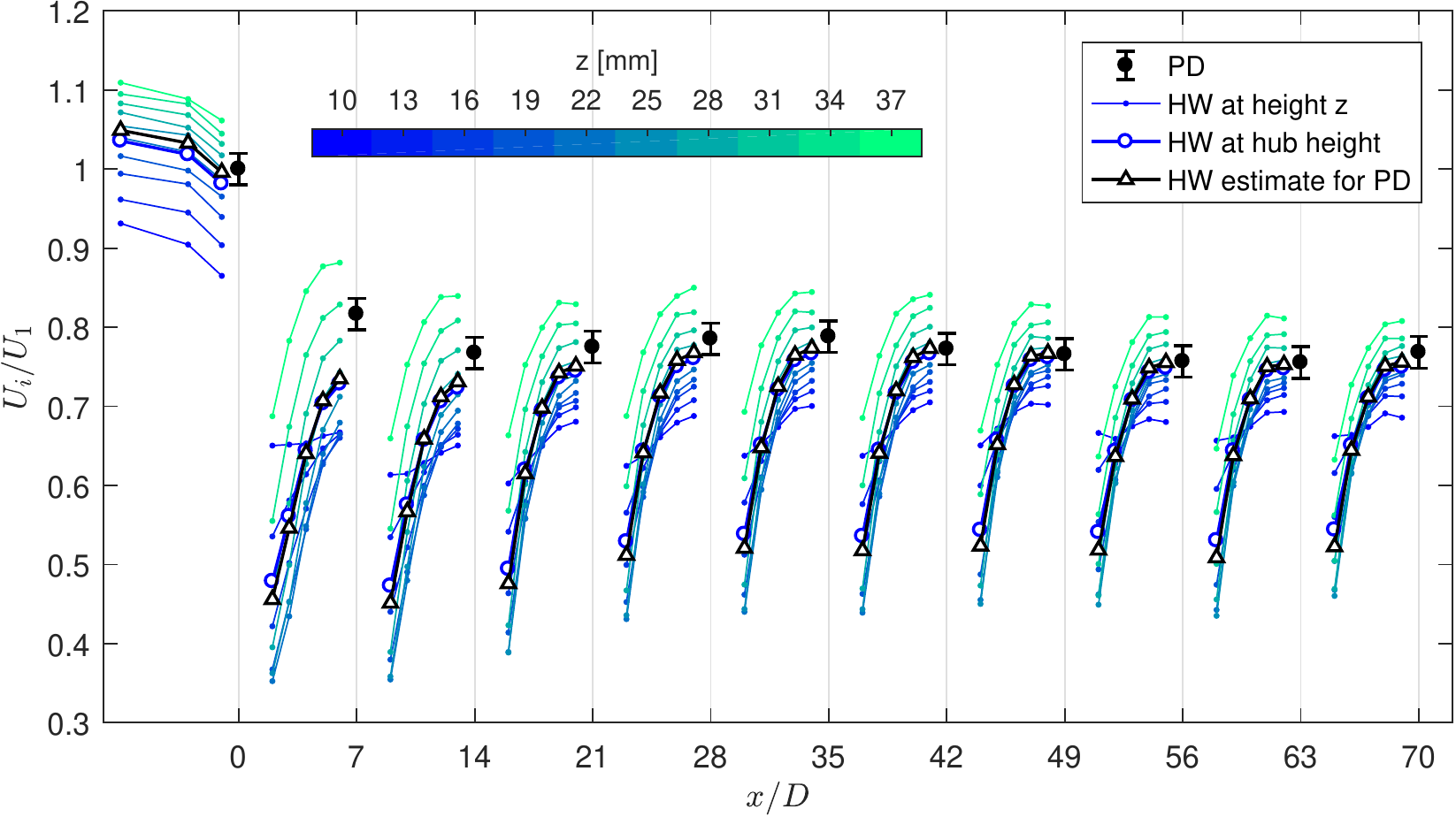}}
		\subfigure[]{\includegraphics[width =0.8 \textwidth, keepaspectratio]{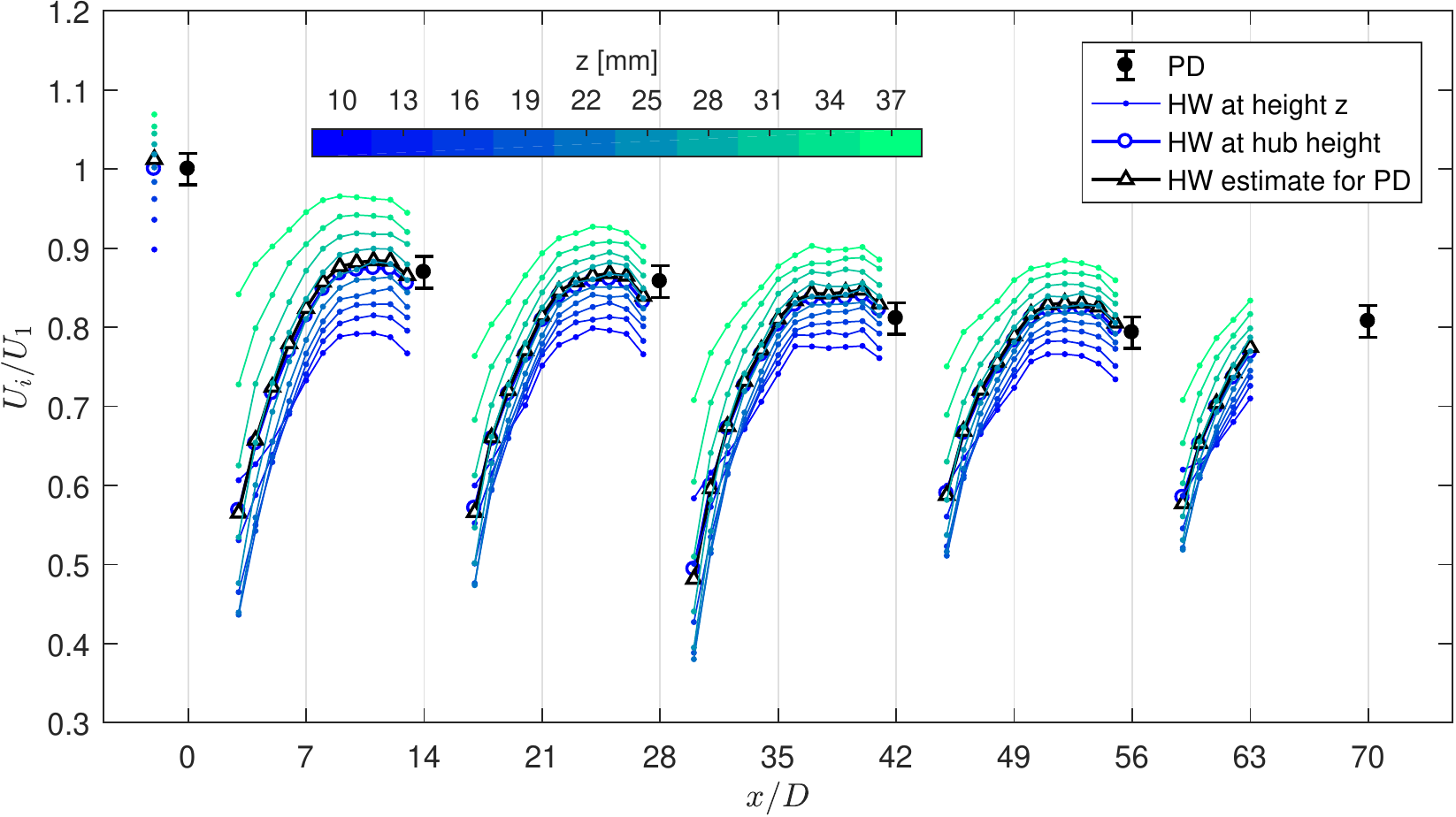}}
	\end{subfigmatrix}
	\caption{Comparison of the hot-wire measurements (HW) of the mean streamwise velocity with the spatially averaged velocity estimated by the porous disk models (PD) for an aligned (a) and staggered (b) layout.}
	\label{hw_comp_U}
\end{figure}

The mean streamwise velocity contours indicate the presence of wakes behind the porous disk models. For the staggered layout, it can be seen how the wakes recover more before they reach the next row, due to the larger streamwise spacing. The contour plots of streamwise turbulence intensity show the highest values in the shear layer at the top-height of the porous disk models. At the bottom of the porous disk models a small peak is observed. The wake is the strongest after the first row. Further downstream, the wake recovery increases thanks to the higher levels of turbulence, caused by the wakes. These results are qualitatively in good agreement with experimental and numerical studies of rotating wind turbine models \cite{Chamorro2011Turbulent,Chamorro2011turbulentb,wu2013simulation}.

\begin{figure}
	\begin{subfigmatrix}{2}
		\subfigure[]{\includegraphics[width = 0.8\textwidth, keepaspectratio]{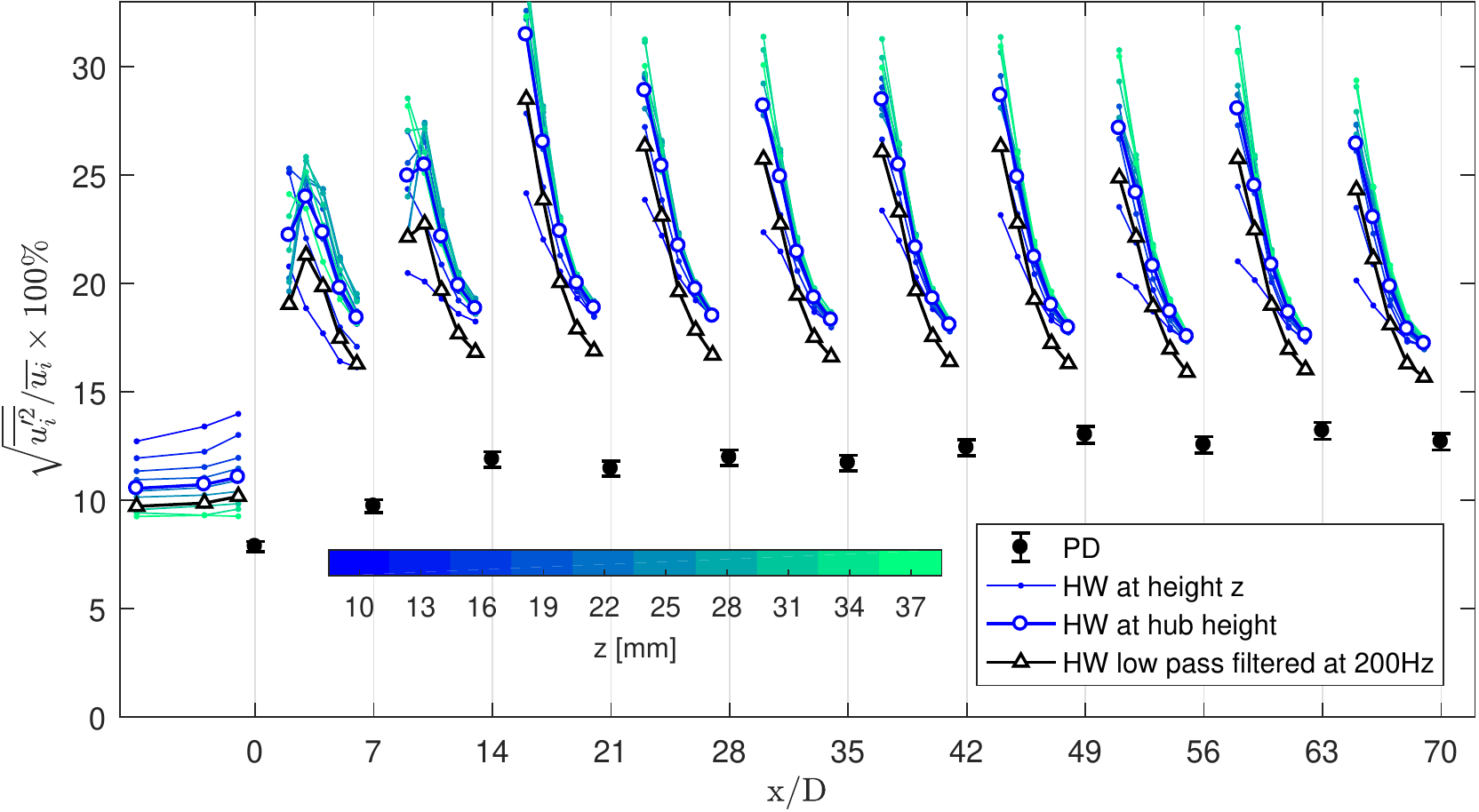}}
		\subfigure[]{\includegraphics[width = 0.8\textwidth, keepaspectratio]{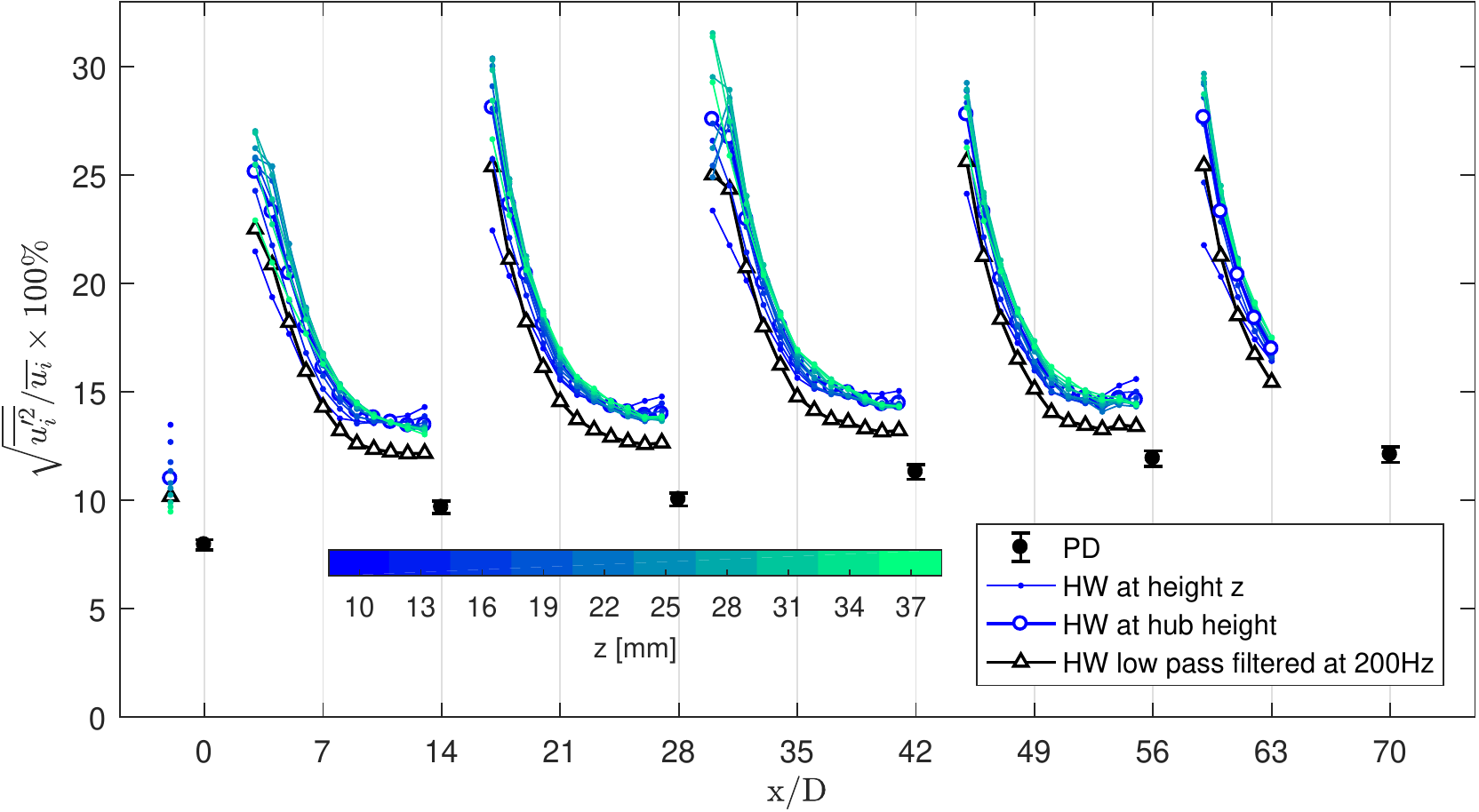}}
	\end{subfigmatrix}
	\caption{Comparison of the local turbulence intensity, measured by the hot-wire probe (HW), and by the porous disk models (PD) for an aligned (a) and staggered (b) layout.}
	\label{hw_comp_TI}
\end{figure}

The hot-wire measurements are compared with the porous disk results in figures \ref{hw_comp_U} and \ref{hw_comp_TI}. Because of the velocity shear in the boundary layer, the spatially averaged measurements by the porous disk models cannot be directly compared to a specific point measurement from the hot-wire probe. All single point hot-wire measurements are shown for a height range that covers $ z_h - D/2 \leq z \leq z_h + D/2$. Here $z_h$ is the hub height of the porous disk model, $D$ the diameter and $R$ the radius. 

The hot-wire velocity is normalized by the velocity measured at hub height and $2D$ upstream of the first wind turbine. For the aligned case, the reference velocity was taken as the average of the measurements at an upstream location of $1D$ and $3D$, as no measurement data was available at a location of $2D$. The porous disk velocities are normalized by the velocity measured by the first model in the farm. The hot-wire measurements visualize the wake recovery. The results for the staggered layouts show a decrease of the velocity in front of each porous disk model, which is not measured for the aligned layout, except for the first row. 

Comparing the the velocity measurements by the porous disk models with the hot-wire probe, a difference is observed for the second and third row, where the porous disk models overestimate the centerline velocity. We expect that a main contributor to this difference is the fact that the porous disk measures the thrust force, which scales with the square of the velocity. From the hot-wire profiles, it can be seen how the shear of mean velocity is larger especially in front of the second and third row, which is expected to play a role in the higher value of the porous disk velocity for those rows. A reconstruction of the measured drag force and corresponding velocity, by making use of the hot-wire profiles, does not fully explain the observed difference. We expect that the measured vertical hot-wire profiles in the center of the porous disk do not provide sufficient information, and that actually the entire cross-plane velocity field in front of the porous disk is necessary to correctly estimate the reconstructed velocity by the porous disk. Considering that all other porous disk models show a much better agreement, it may be possible that the measurements by the porous disks in row 2 are also influenced by an unexpected measurement error. For all other porous disk models, the reconstructed velocity measurements match the hot-wire results at hub height very well, for both the aligned and staggered layout, confirming the measurement capabilities of the setup in general.

Figure \ref{hw_comp_TI} shows the local turbulence intensity measured by the porous disks and the hot-wire probe. The local turbulence intensity is based on the local velocity of the hot-wire probe, or the spatially averaged velocity from the porous disk. The signal measured by the porous disk is filtered twice, once by a digital low-pass filter (a digital sharp cut-off filter at $200\,{\rm Hz}$ is applied in the post-processing) and once due to spatial averaging over the disk. The porous disk thus measures lower turbulence levels then the hot-wire probe. For comparison, figure \ref{hw_comp_TI} also shows the turbulence intensity calculated from the hot-wire velocity, after applying a similar sharp cut-off filter at $200\,{\rm Hz}$. The turbulence intensity after filtering the hot-wire signals, shown with the black lines, are only slightly lower than the unfiltered levels, indicating that most of the energy-containing fluctuations are found below $200\,{\rm Hz}$. The largest part of the spectral filtering for the porous disk is thus a result of the spatial averaging over the disk. The actual amount of filtering by the porous disk depends on the original spectrum of the velocity fluctuations, and thus varies from row to row.

\section{Wind farm measurements}
\label{s:wf_measurements}

In this section the wind farm measurement results are presented. First the layouts with a uniform streamwise spacing are discussed. Then the benefits of a moderate (\textit{NU1}), and a more extreme (\textit{NU2}) non-uniform streamwise spacing are presented.

\subsection{Uniform spacing}

\begin{figure}
	\begin{subfigmatrix}{2}
		\subfigure[]{\includegraphics[width = 0.7\textwidth, keepaspectratio]{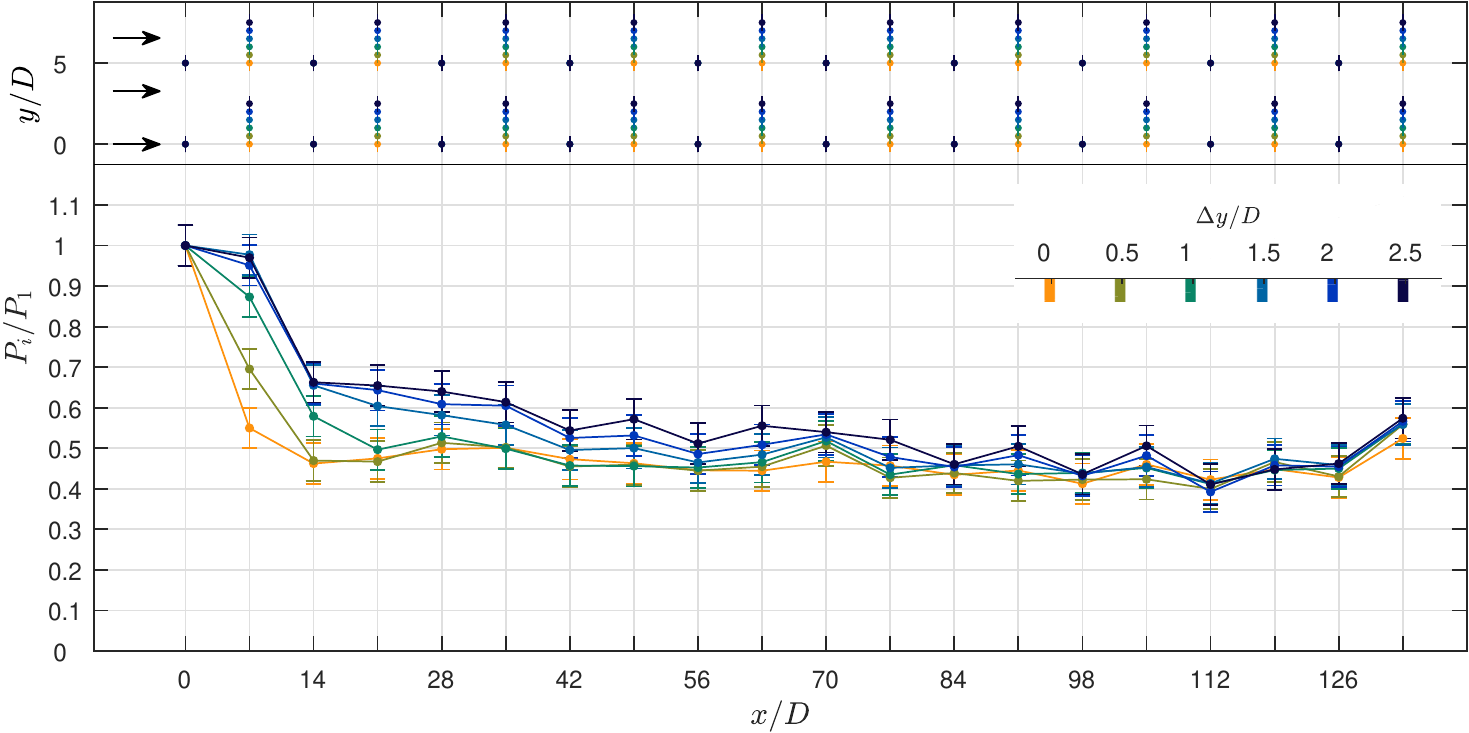}}
		\subfigure[]{\includegraphics[width = 0.7\textwidth, keepaspectratio]{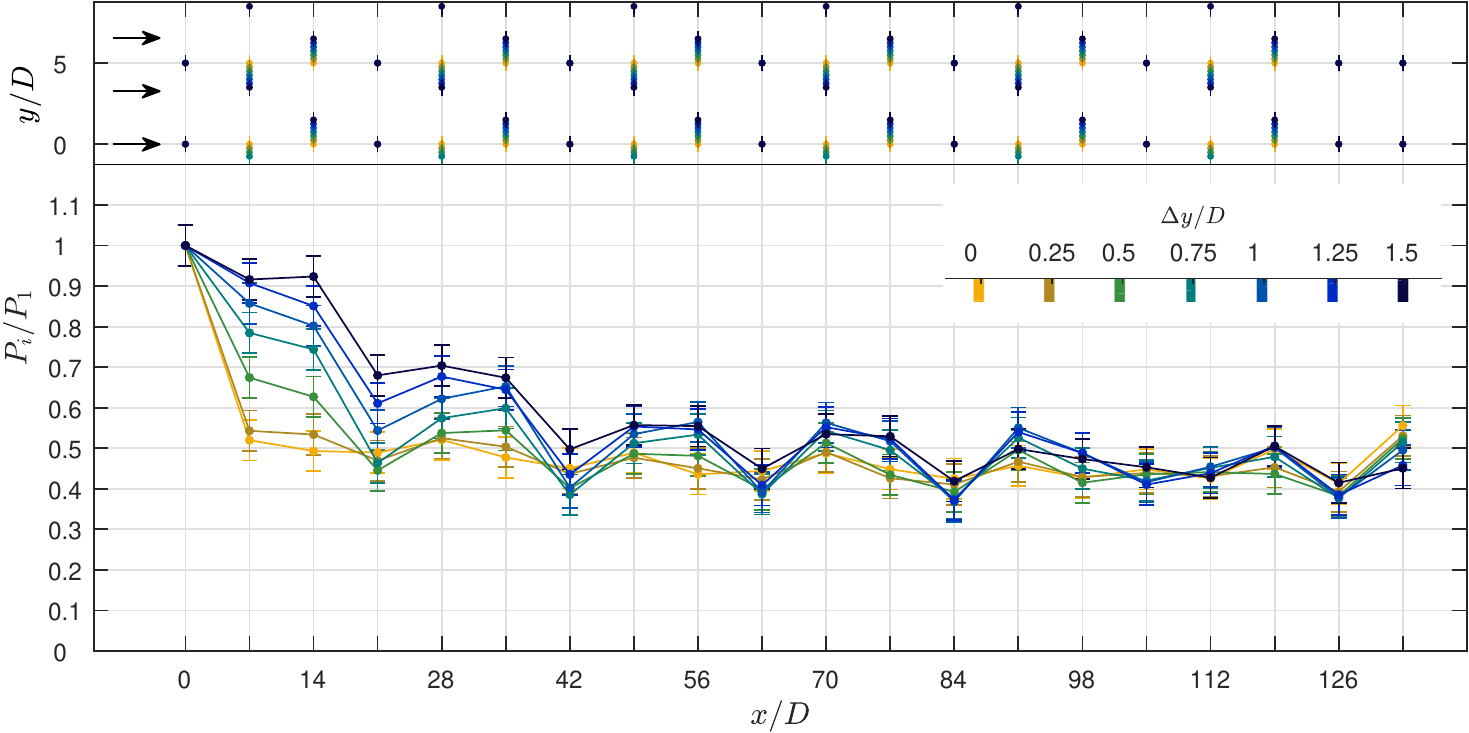}}
	\end{subfigmatrix}
	\caption{Wind farm measurements of the mean surrogate power in each row for (a) the \textit{U-C1} and (b) \textit{U-C2} layout series. The layouts are indicated with corresponding colors on top of each figure. See figure \ref{f:schematic_wt} for an overview of all layouts.}
	\label{f:WF_U}
\end{figure}

\begin{figure}
	\begin{subfigmatrix}{2}
		\subfigure[]{\includegraphics[width = 0.7\textwidth, keepaspectratio]{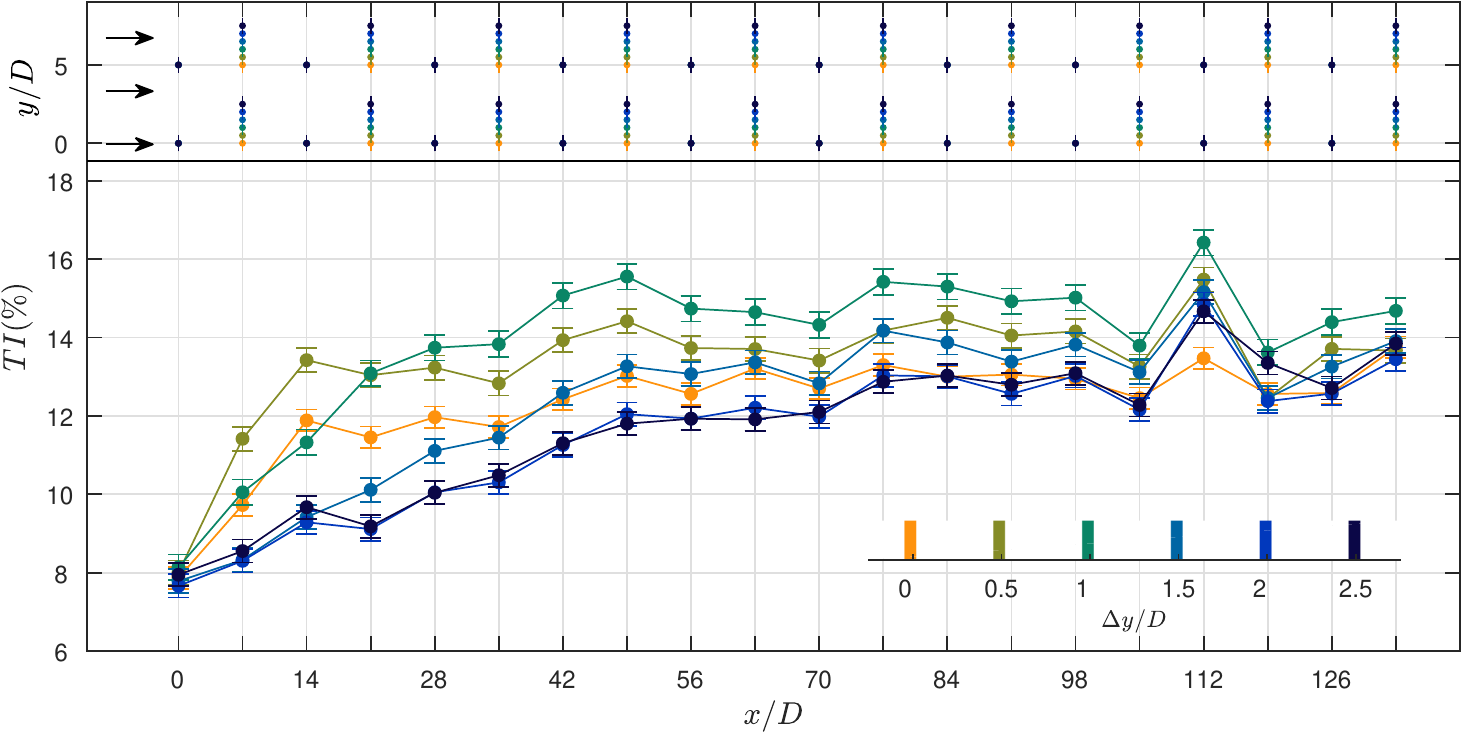}}
		\subfigure[]{\includegraphics[width = 0.7\textwidth, keepaspectratio]{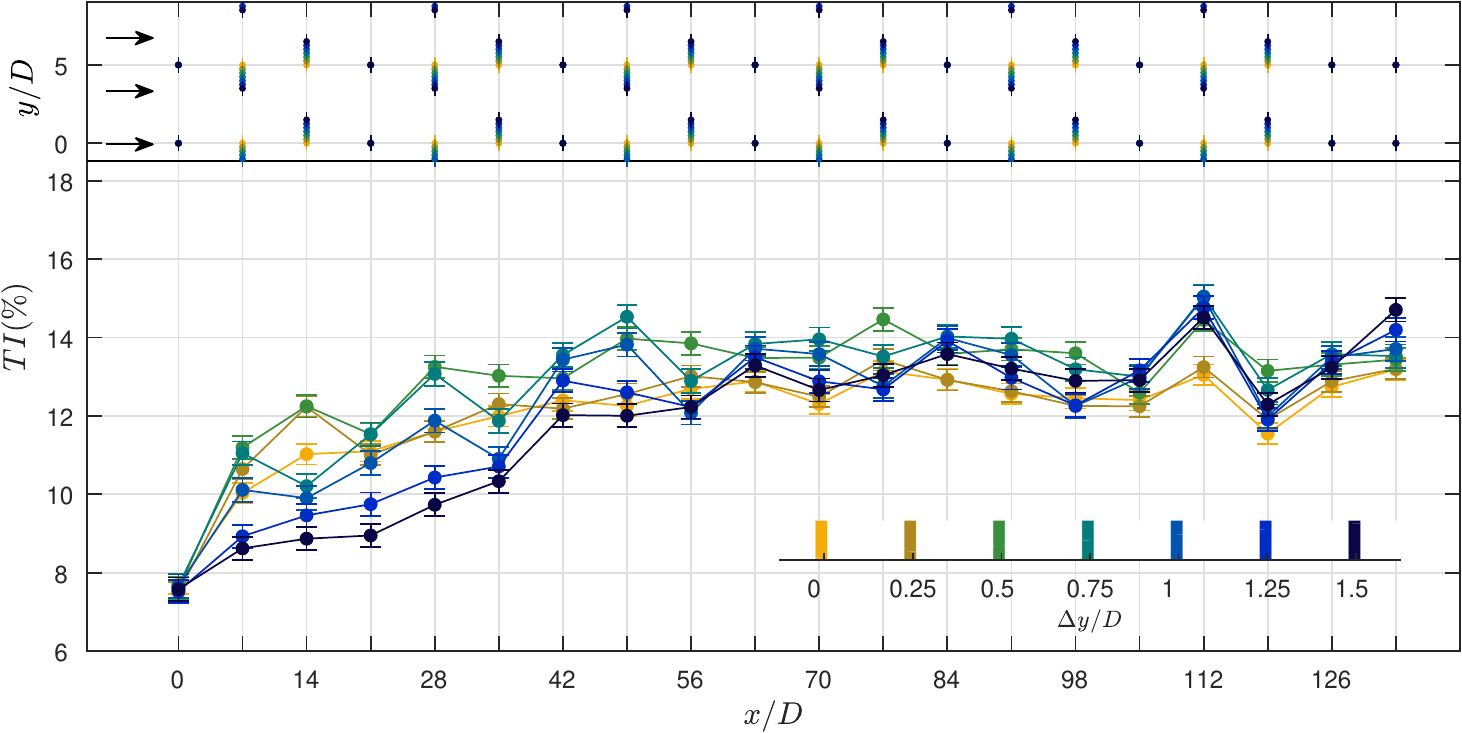}}
	\end{subfigmatrix}
	\caption{Turbulence intensity as measured by the porous disk models in each row for (a) the \textit{U-C1} and (b) the \textit{U-C2} layout series. The layouts are indicated with corresponding colors on top of each figure. See figure \ref{f:schematic_wt} for an overview of all layouts.}
	\label{f:WF_U_TI}
\end{figure}

Figures \ref{f:WF_U} and \ref{f:WF_U_TI} present the results for the first two layout series, \textit{U-C1} and \textit{U-C2}. The first series represents the change of a regular array, from aligned to staggered. When the layout is fully aligned, the mean row power reduces quickly over the first three rows, after which it levels off and slowly reduces from $P_i/P_1 \approx 0.45$ , to $P_i/P_1 \approx 0.4$ over the next fifteen rows. Considering the differences in boundary layer conditions, these power losses are the same order of magnitude as the losses of $45\%$ in the Horns Rev wind farm \cite{barthelmie2011flow} with a similar streamwise spacing, almost $50\%$ in the Walney 2 wind farm with also a similar streamwise spacing \cite{nygaard2014wakes}, or more than $60\%$ observed in the Middelgrunden offshore wind farm \cite{barthelmie2007modelling} with a smaller spacing. When the layout is changed from aligned to staggered, the surrogate power increases mainly for the first ten to fifteen rows, indicating a slower move towards a fully-developed regime. Interestingly, at the end of the wind farm, little differences are seen compared to the aligned configuration: both layouts tend to the same asymptotic limit. The staggered layout results in the highest total farm surrogate power output as a result of its higher power output in the entrance region. Furthermore, when staggered, the first two rows measure approximately the same surrogate power and turbulence intensity, indicating that the second row sees approximately an unperturbed free stream flow.

For every layout, it is noticed that the last row consistently measures a higher surrogate power. It is possible that this offset is related to its location, very close to the end of the wind tunnel test-section, where the test section has a slight contraction. This argument is supported by the observation that for the layout \textit{NU2-C1}, where the last row is shifted $5.5D$ upstream, the effect of a higher mean power increase for the last row is reduced significantly. To exclude this effect from the analysis, we do not include the last row when we study the asymptotic behavior in section \ref{s:wf_discussion}.

The mean power for the \textit{U-C2} layouts show the same trends. By shifting the rows to a double staggered configuration, the surrogate power increases mainly in the first ten-to-fifteen rows. However, the increase in the first half of the wind farm is larger than before. Within the measurement uncertainty, it is possible to recognize a pattern for each three consecutive rows, as a consequence of the repeating layout. The second and third row of the wind farm, show almost the same surrogate power as the first row. Further downstream, it is the rows that are not moved, i.e. the first row in each pattern of three \mbox{(row 4, 7, 10, 13, ..)}, that show a lower surrogate power, or larger wake losses.

Because it is impossible to accommodate the three-row pattern until the end of the twenty-row wind farm, the last two rows were kept unchanged in the aligned configuration. This explains the lower power for the last two rows. In the last part of the wind farm, the pattern is also more difficult to distinguish, which could be a result of the measurement uncertainty. Qualitatively, both layouts, \textit{U-C1} and \textit{U-C2}, tend to the same asymptotic limit in the fully developped regime. The reconstructed turbulence intensity (indicative for unsteady loading) as measured for \textit{U-C1} and \textit{U-C2} is shown in figure \ref{f:WF_U_TI}. For the aligned layout, the turbulence intensity increases fast in the first three rows, and eventually levels off after about twelve rows. This trend indicates that while the power levels off quickly in the first few rows, the flow is still developing until further into the wind farm (in this case approximately the twelfth row), as also observed by Ref. \cite{Chamorro2011Turbulent}. The staggered layout results in a smaller unsteady loading, which increases more slowly with row number, but eventually reaches the same level as the aligned layout at the end of the wind farm, e.g. $TI \approx  13 \%$. It is interesting to note that while all layouts tend to the same mean power asymptote, the unsteady loading shows different asymptotes, with higher values for \textit{U-C1} layouts with a spanwise shift smaller than $1D$. In these cases the porous disk models have a partial wake overlap which is expected to cause the higher variability. These slightly-shifted layouts are thus not preferred, as they result in a below-optimal power output and the highest level of unsteady loading.

The \textit{U-C2} layout-series shows similar trends for the unsteady loading. The double staggered layout results in a similar slow increase, with at the end also a turbulence intensity of $TI \approx  13 \%$. In this case, the intermediate layouts only result in a slightly higher unsteady loading, thanks to the increased streamwise spacing of a double staggered approach.

\subsection{Moderate non-uniform spacing}

\begin{figure}
	\begin{subfigmatrix}{2}
		\subfigure[]{\includegraphics[width = 0.7\textwidth, keepaspectratio]{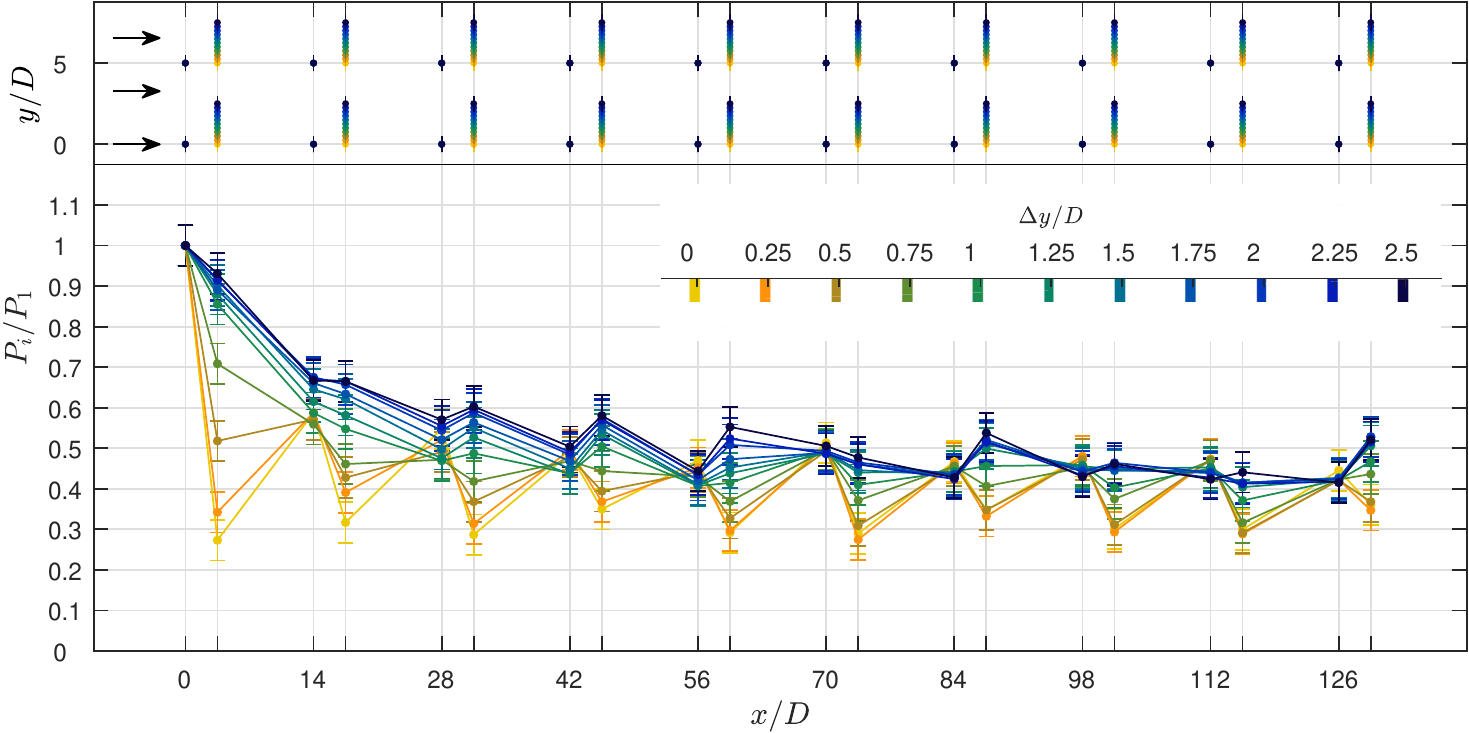}}
		\subfigure[]{\includegraphics[width = 0.7\textwidth, keepaspectratio]{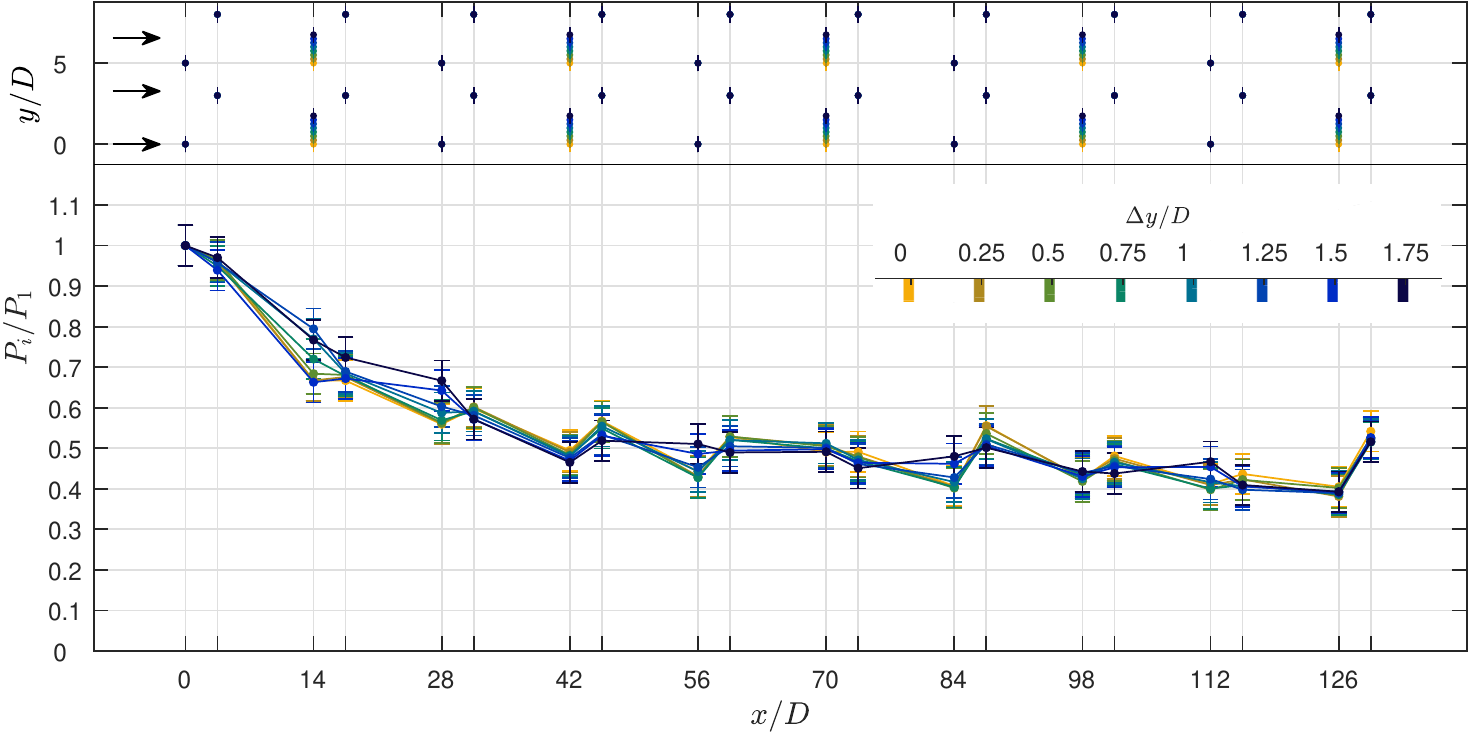}}
	\end{subfigmatrix}
	\caption{Wind farm measurements of the mean surrogate power in each row for (a) the \textit{NU1-C1} and (b) \textit{NU1-C2} layout series. The layouts are indicated with corresponding colors on top of each figure. See figure \ref{f:schematic_wt} for an overview of all layouts.}
	\label{f:WF_NU1}
\end{figure}

\begin{figure}
	\begin{subfigmatrix}{2}
		\subfigure[]{\includegraphics[width = 0.7\textwidth, keepaspectratio]{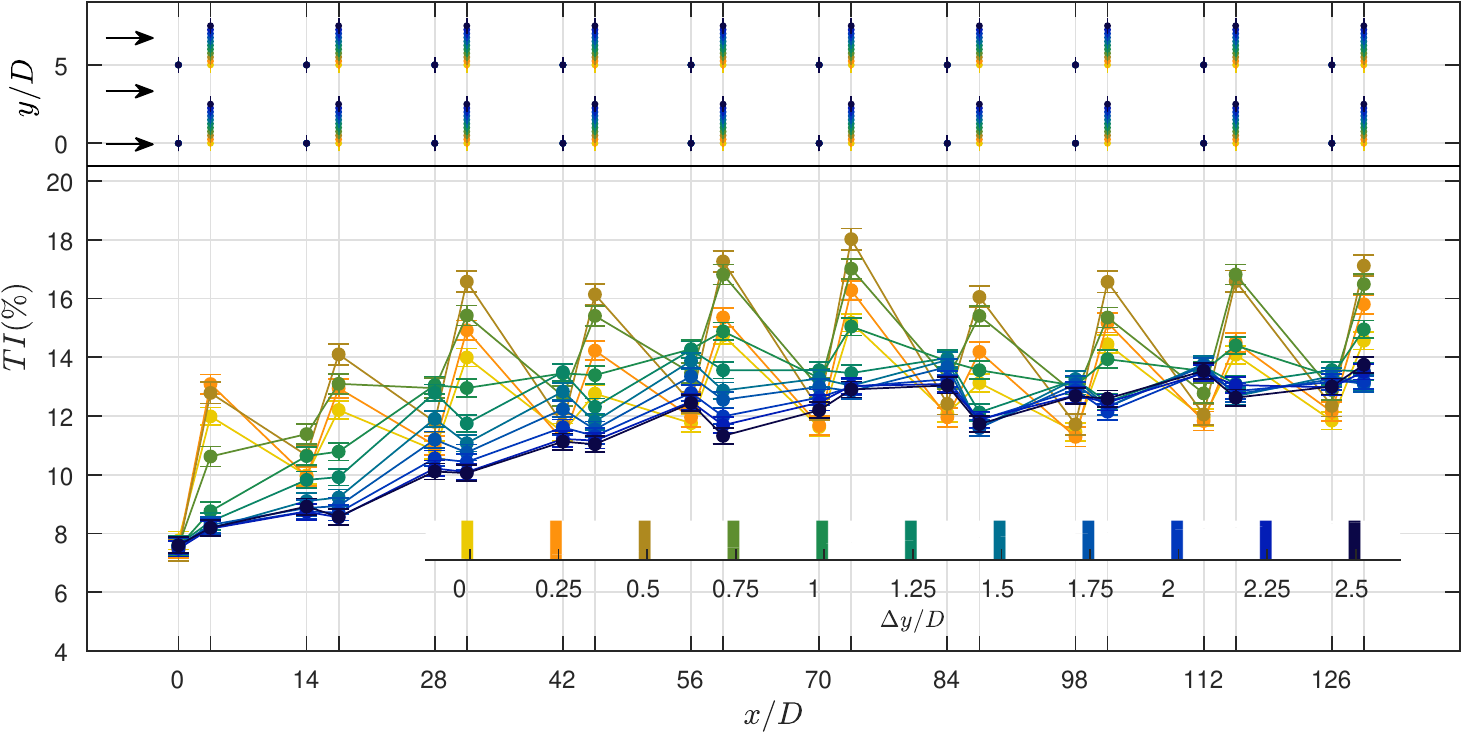}}
		\subfigure[]{\includegraphics[width = 0.7\textwidth, keepaspectratio]{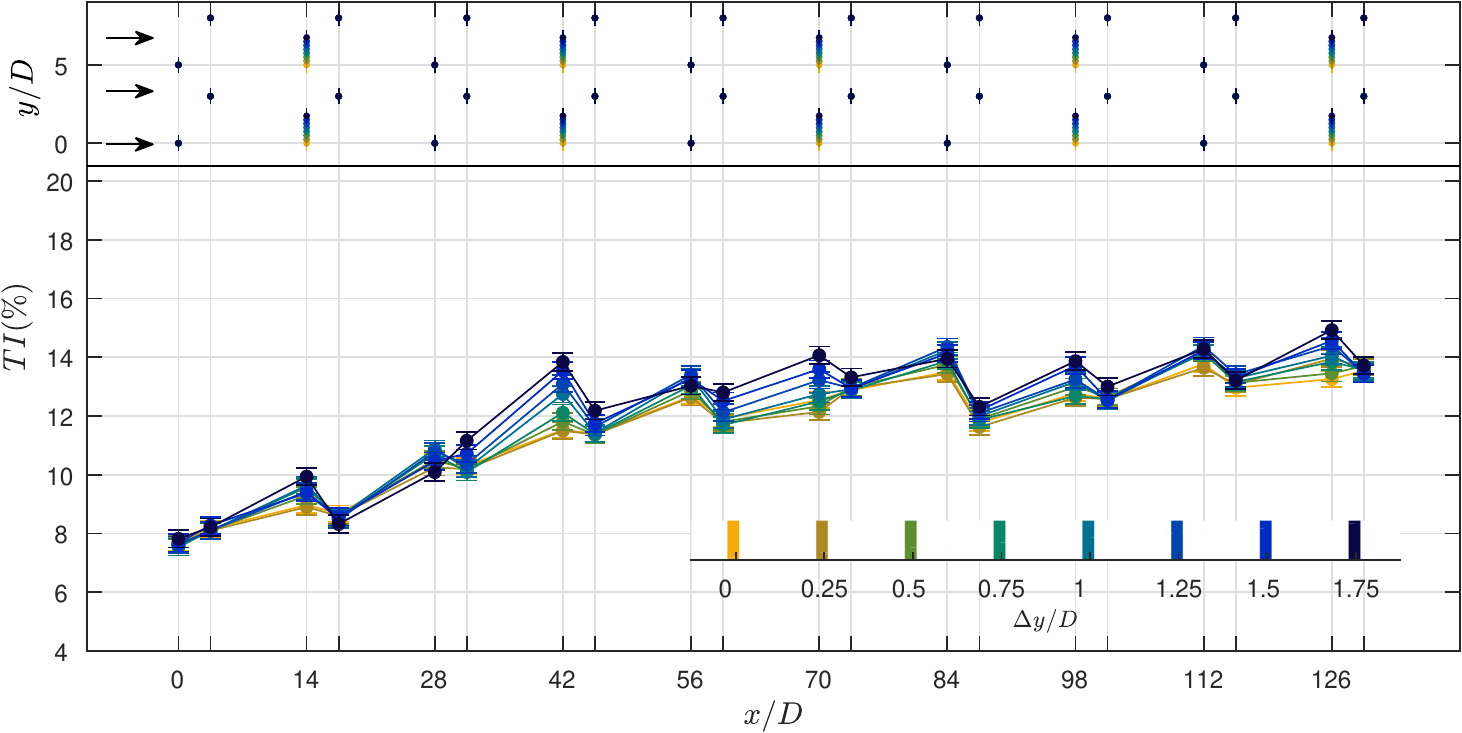}}
	\end{subfigmatrix}
	\caption{Turbulence intensity as measured by the porous disk models in each row for (a) the \textit{NU1-C1} and (b) \textit{NU1-C2} layout series. The layouts are indicated with corresponding colors on top of each figure. See figure \ref{f:schematic_wt} for an overview of all layouts.}
	\label{f:WF_NU1_TI}
\end{figure}

The measurement results for the non-uniform layouts series \textit{NU1} are shown in \mbox{figures \ref{f:WF_NU1}} \mbox{and \ref{f:WF_NU1_TI}}. For an aligned configuration, the disadvantage of smaller turbine distances (i.e. $ S_x/D = 3.5$ instead of $s_x/D = 7$) is clear: every second row shows a very low surrogate power output and high unsteady loading, associated with their location in the near wake from an upstream model. The rows with a larger upstream streamwise spacing (i.e. $ S_x/D = 10.5$ instead of $s_x/D = 7$) do measure a higher power, e.g. $P_3/P_1 \approx 0.6$ compared to $P_3/P_1 \approx 0.45$ for the original aligned layout. However, these improvements do not compensate the significantly lower outputs of the closely spaced models.

By sliding the even rows in the spanwise direction, the impact of wakes is reduced significantly. Most of the improvements are made by shifting from $0D$ to $1D$. Increasing the spanwise shift of the even rows to a fully staggered layout results in the highest surrogate power output. The mean row power for the staggered configuration follows a very similar trend as the previous results for a uniformly spaced staggered wind farm. The surrogate power is the highest at the beginning of the wind farm, and reduces towards an asymptote at the end. Interestingly, the staggered layout shows a repeating pattern for each pair of consecutive rows. The even rows (starting from row 6) which are closely spaced and staggered with the upstream uneven rows, measure a higher surrogate power, which indicates less wake losses, or possibly the presence of a local flow interaction, similar to observed by \cite{mctavish2014experimental}. However, a clear trend is not obvious. As before, the fully staggered layout results in the lowest unsteady loading. The turbulence intensity levels off after approximately 13 rows, reaching a value of $TI \approx  13 \%$, similar to the observation for the previous layout series. 

The measurements for the \textit{NU1-C2} series show no clear benefits for the power. While the second to fifth row increase for the largest spanwise shift, the power decreases slightly everywhere else in the wind farm. Interestingly, also the unsteady loading of the porous disk models increases with increasing spanwise shift. It is concluded that the \textit{NU1-C2} layout series brings no direct benefits for power output or unsteady loading.

\begin{figure}
	\begin{subfigmatrix}{3}
	\subfigure[]{\includegraphics[width = 0.7\textwidth, keepaspectratio]{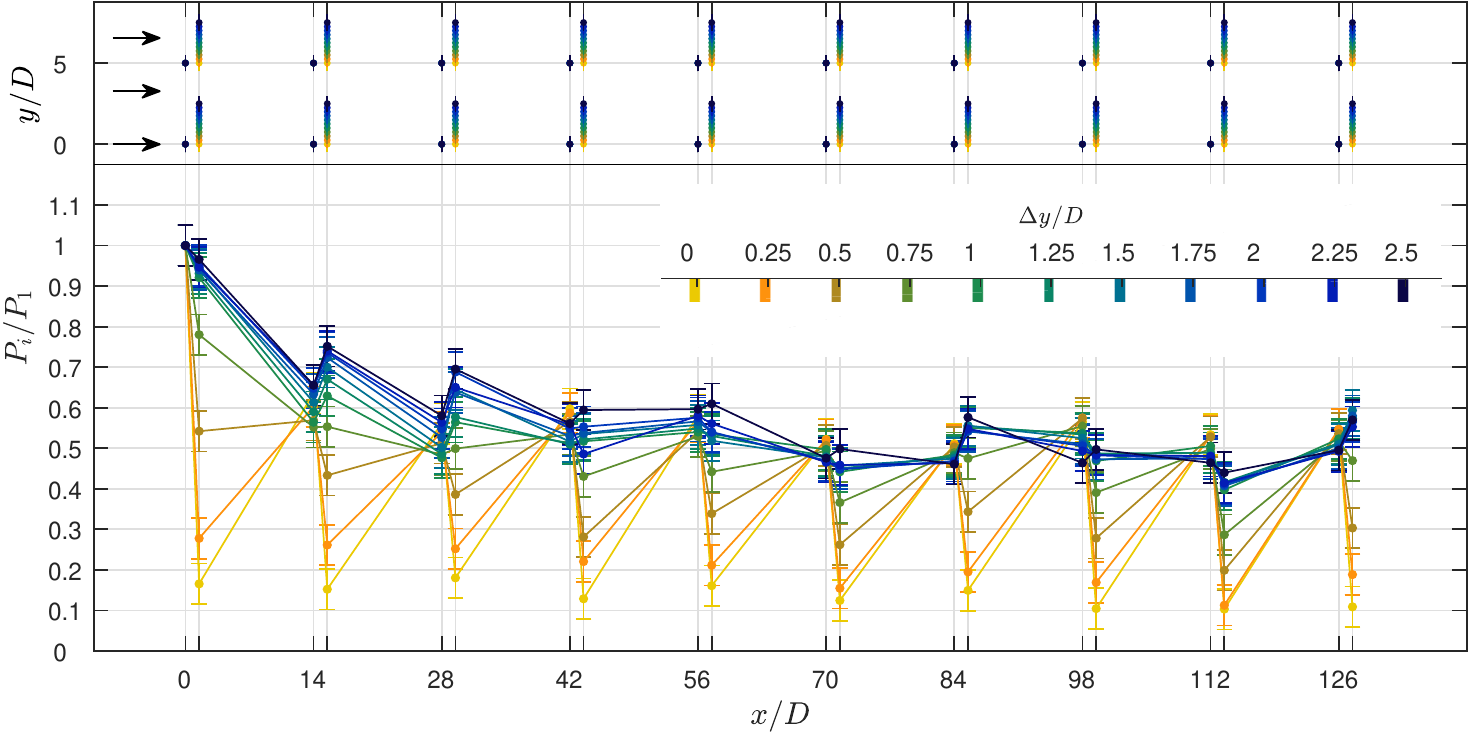}}
	\subfigure[]{\includegraphics[width = 0.7\textwidth, keepaspectratio]{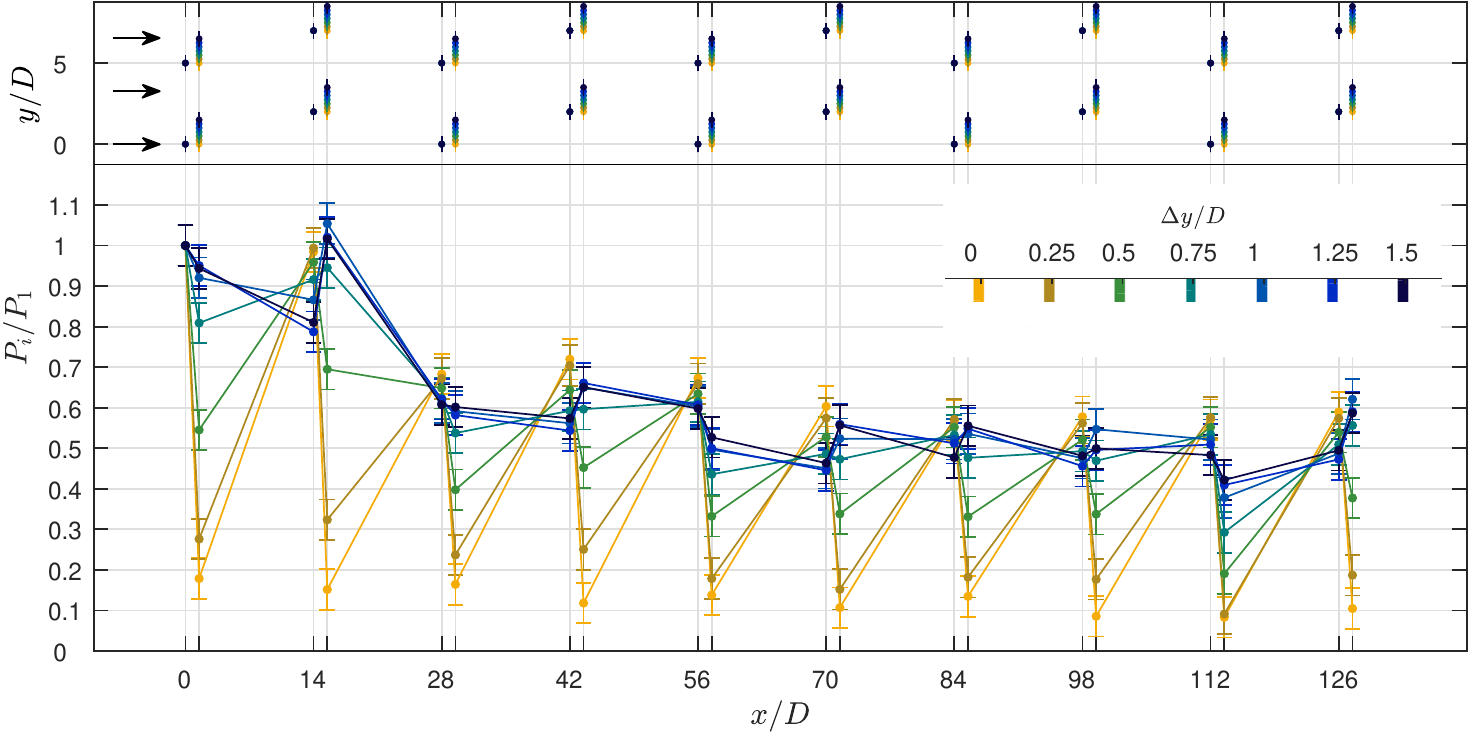}}
	\subfigure[]{\includegraphics[width = 0.7\textwidth, keepaspectratio]{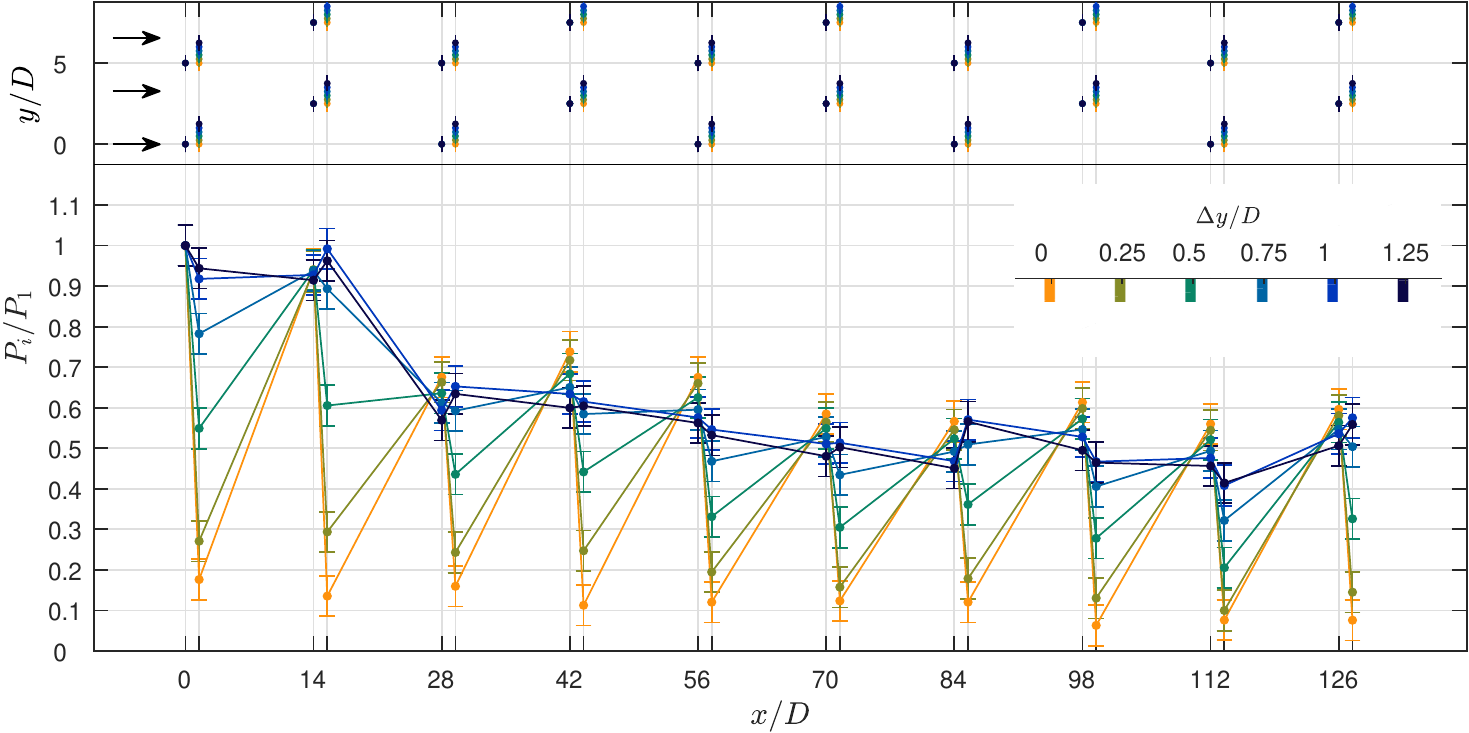}}
	\end{subfigmatrix}
	\caption{Porous disk measurements of mean surrogate power in each row for (a) the \textit{NU2-C1}, (b) the \textit{NU2-C2}, and (c) the \textit{NU2-C3} layout series. The layouts are indicated with corresponding colors on top of each figure. See figure \ref{f:schematic_wt} for an overview of all layouts.}
	\label{f:WFNU2_P}
\end{figure}

\subsection{Extreme non-uniform spacing}

\begin{figure}
	\begin{subfigmatrix}{3}
		\subfigure[]{\includegraphics[width = 0.7\textwidth, keepaspectratio]{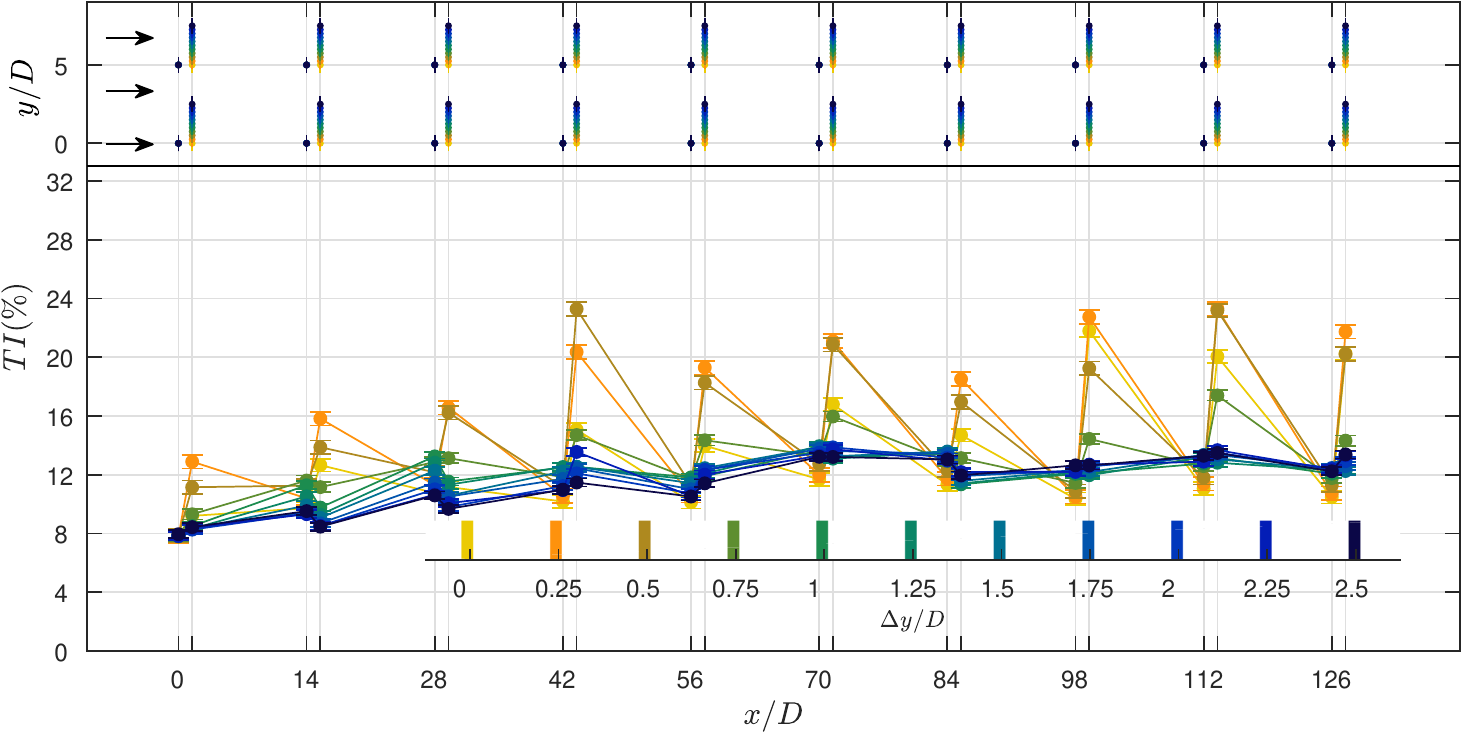}}
		\subfigure[]{\includegraphics[width = 0.7\textwidth, keepaspectratio]{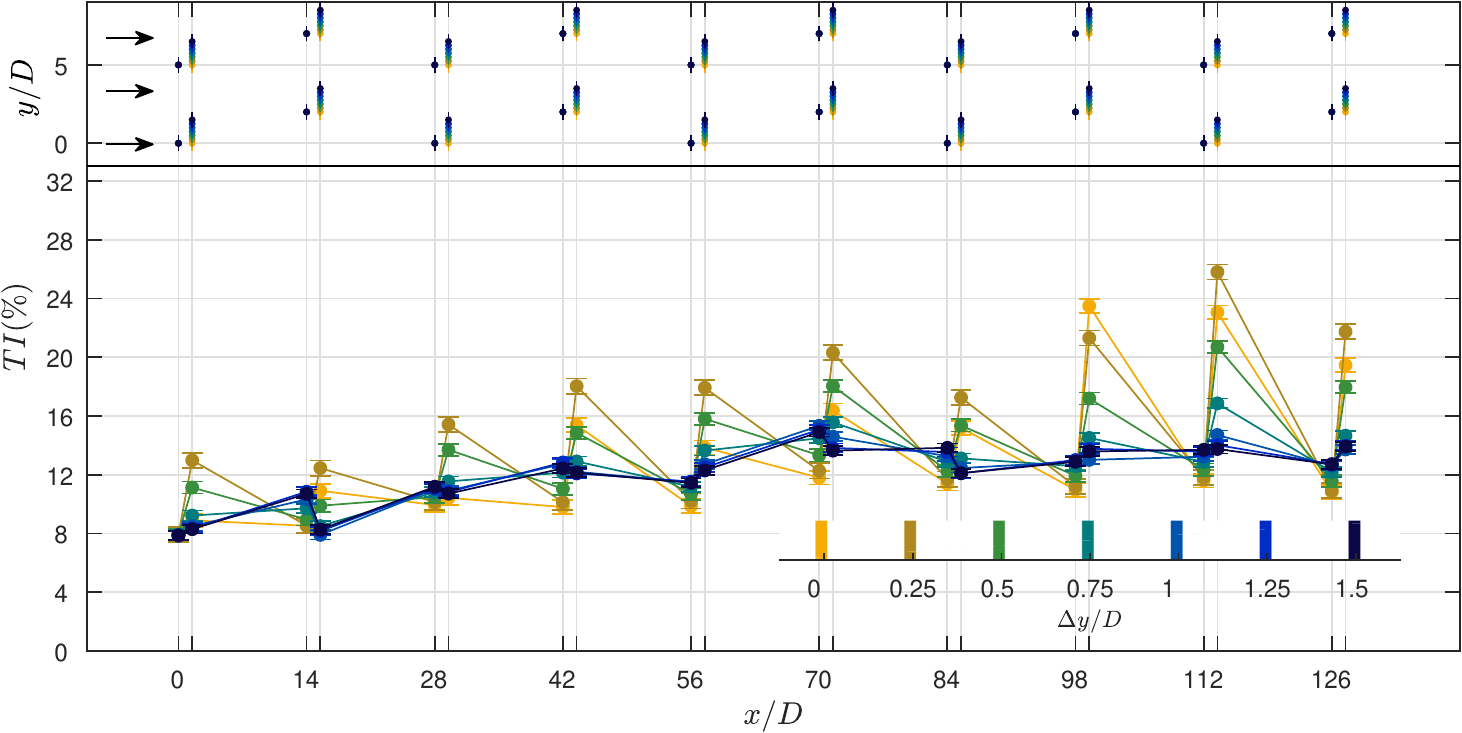}}
		\subfigure[]{\includegraphics[width = 0.7\textwidth, keepaspectratio]{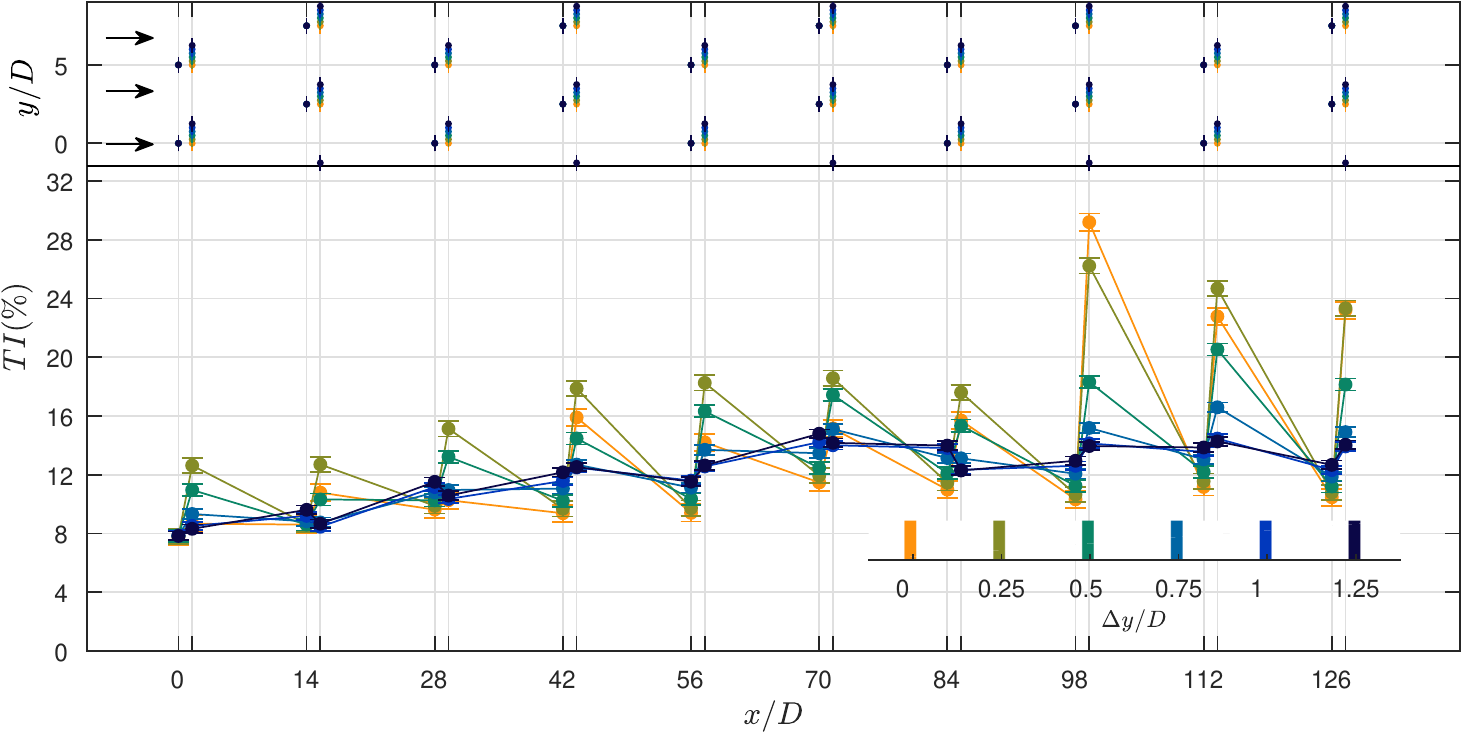}}
	\end{subfigmatrix}
	\caption{Porous disk measurements of turbulence intensity in each row for (a) the \textit{NU2-C1},(b) the  \textit{NU2-C2}, and (c) the \textit{NU2-C3} layout series. See figure \ref{f:schematic_wt} for an overview of the layouts.}
	\label{f:WFNU2_TI}
\end{figure}

The measured surrogate power output for the \textit{NU2} series are shown in figure \ref{f:WFNU2_P}, and the estimated turbulence intensity is shown in figure \ref{f:WFNU2_TI}. This layout series pursues an extremely uneven streamwise spacing. As a result, the even rows in the aligned configurations measure a very low surrogate power output, of approximately $P_i/P_1 \approx 0.1 - 0.2$.

The \textit{NU2-C1} layout series shows similar trends as the \textit{NU1-C1} series, however, with a better performance in the staggered configuration. For this layout, every even row measures the same or higher power than the upstream row, indicating less wake losses or a possible local flow interaction, e.g. the local blockage results in a slight acceleration towards the downstream model similar to observations by Ref. \cite{mctavish2014experimental}. Qualitatively, the mean row power reduces less quickly, with row $10$ measuring a surrogate power output of $P_i/P_1 \approx 0.6$.

For the staggered \textit{NU2-C2} layout, the power of the first four rows does not drop significantly, and the power of the fourth row is approximately equal, or even higher, than the value of the first row (it is important to note that considering the measurement uncertainty the small increase is not statistically significant). Similar to the \textit{NU2-C1} series, every fourth row of each recurring four-row-pattern, displays a slightly higher surrogate power. These observations indicate a possible local acceleration of the flow towards each fourth row. The \textit{NU2-C3} series shows similar trends, however, now the values for each fourth row are slightly lower, while the power of each third row has increased. As a result, the mean row power follows a smoother progression towards an asymptote at the end of the farm. With the layout \textit{NU2-C2} and \textit{NU2-C3} it is thus possible to significantly increase the power of the first four rows, to almost the same value of the first row.

The measurements of local turbulence intensity are shown in figure \ref{f:WFNU2_TI}. When the layouts are aligned, the even rows measure very high values of the local turbulence intensity due to the low velocities in the near wake. However, when the layouts are staggered, a relatively smooth progression is observed, very similar to the other layout series. After about $11$ rows, the local turbulence intensity plateaus to a value of approximately  $TI \approx  13 \%$.

\section{Discussion: wind farm layout}
\label{s:wf_discussion}

The wind farm results in the previous section displayed a number of interesting trends. First, when considering various arrangements, most of the increase in surrogate power output occurs at the beginning of the farm. This observed trend is in good agreement with results in the literature \cite{barthelmie2011flow,stevens2014large}, and indicates the importance of reducing wake losses in the entrance region of the farm. Second, for each series, the layouts with the highest surrogate power show a relatively smooth decrease of the power towards a constant value, or asymptote, at the end of the farm, indicating the approach of a fully-developed flow regime. In this section the entrance and fully-developed region are analyzed as a function of layout by studying the average power of both the whole farm, and of the asymptotic trend as deduced from the last few rows.

\begin{figure}
	\begin{subfigmatrix}{2}
		\subfigure[]{\includegraphics[width = 0.7\textwidth, keepaspectratio]{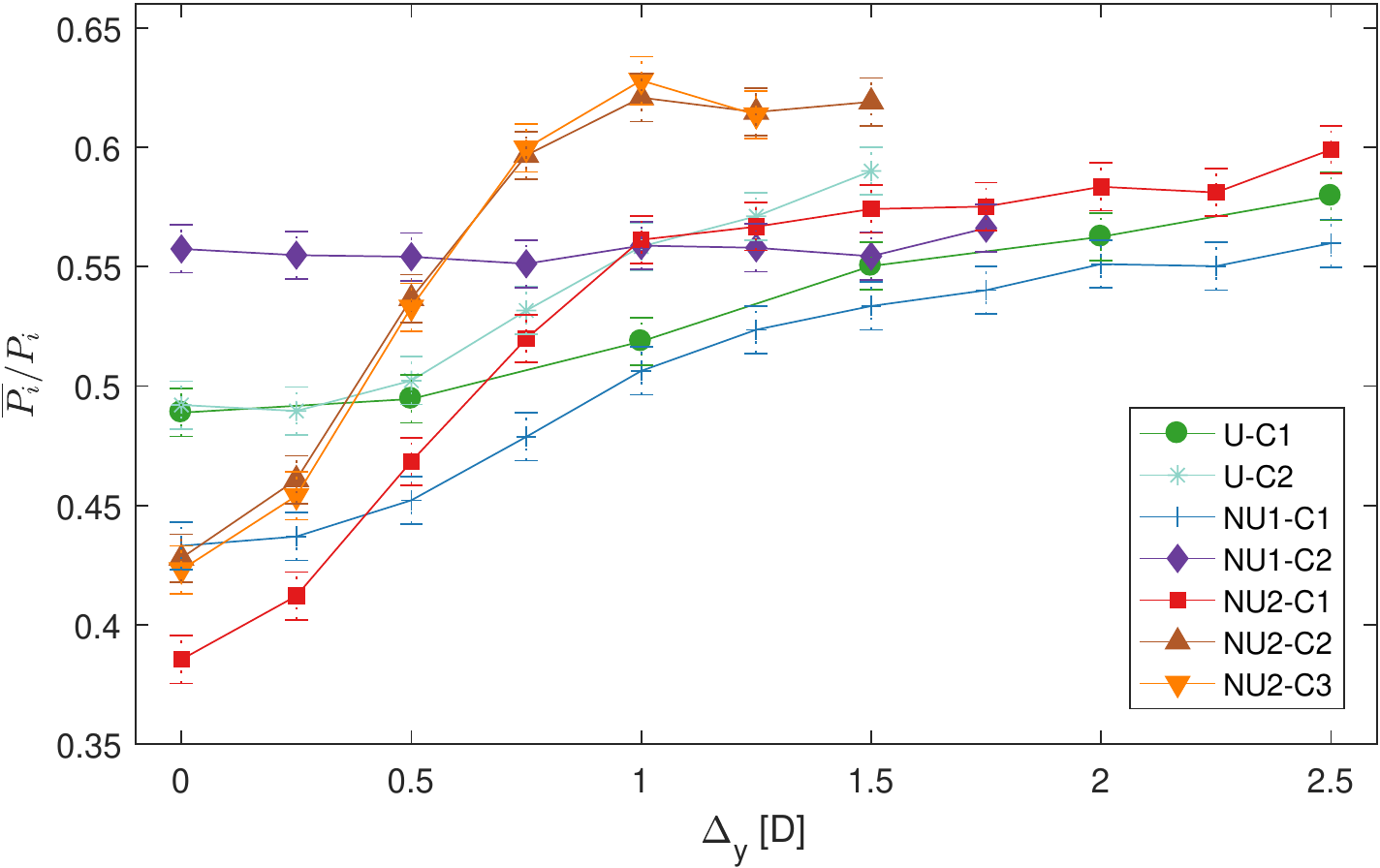}}
		\subfigure[]{\includegraphics[width = 0.7\textwidth, keepaspectratio]{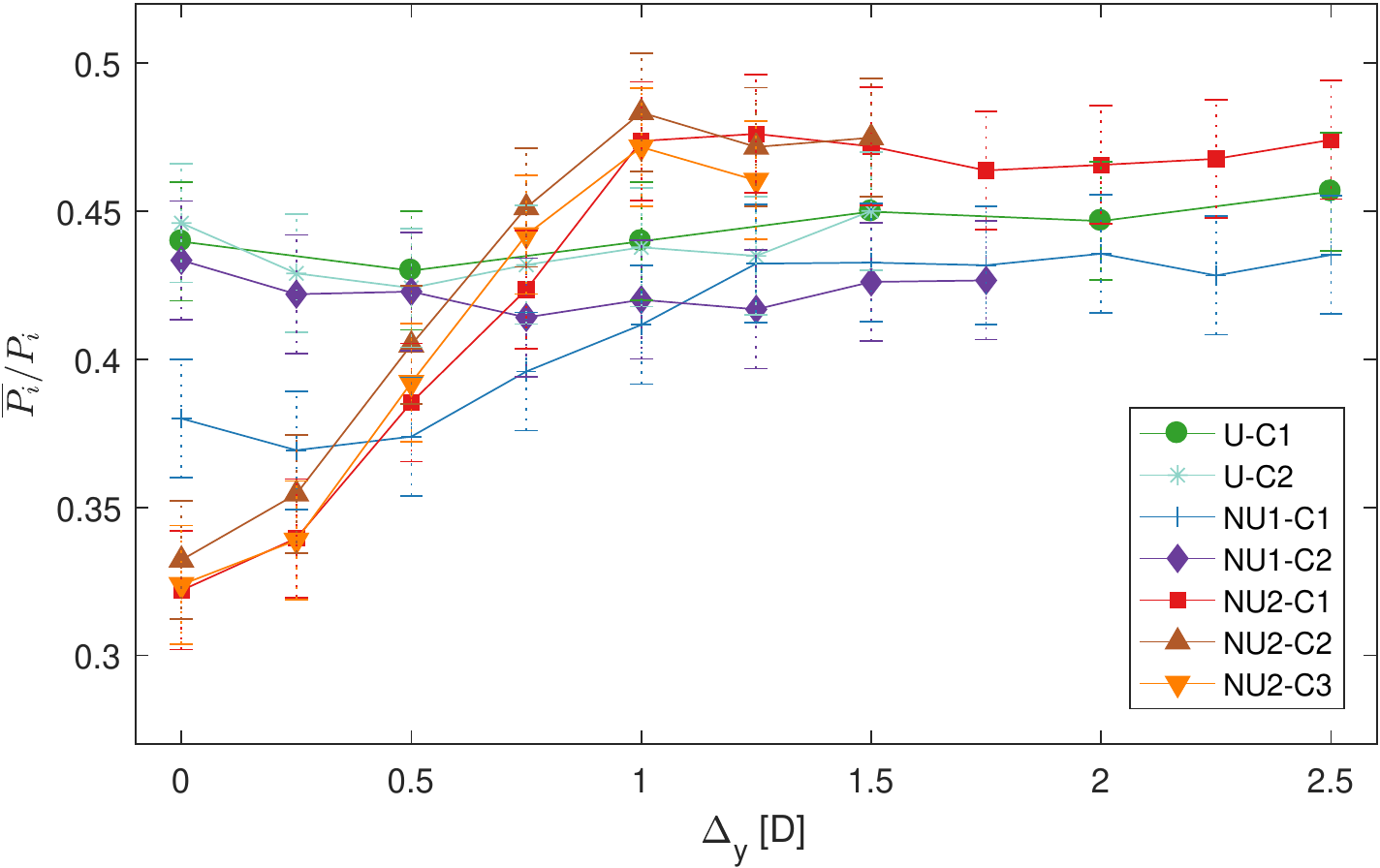}}
	\end{subfigmatrix}
	\caption{ The farm-average surrogate power (\textit{a}) and the average over row 16-19 (\textit{b}) as a function of the spanwise shift $\Delta_y$. See figure \ref{f:schematic_wt} for an overview of the layouts.}
	\label{f:WFAS}
\end{figure}

The farm-averaged surrogate power $\overline{P_i}/P_1 = (1/N)\sum_{i=1}^N P_i/P_1$, where N is the number of porous disk models considered in the aggregate, is shown in figure \ref{f:WFAS} (a) as a function of the spanwise shift $\Delta_y$. If the farm efficiency is defined as the total power output per square area, finding the layout with the highest farm efficiency, is similar to finding the layout with the highest farm-averaged surrogate power $\overline{P_i}/P_1$, since the farm area is a constant in the experiments.

In general, as expected, the lowest farm efficiencies are obtained for a zero spanwise shift, i.e. for aligned cases. The wake losses are especially large for the layouts with an uneven streamwise spacing, as half of the models are spaced very closely (e.g. $1.5D$ for \textit{NU2} and $3.5D$ for \textit{NU1}). The \textit{NU2-C1} series has the lowest efficiency for a zero shift, while the variations \textit{NU2-C2} and \textit{NU2-C3} have a slightly higher efficiency. 

From the first two layout series with a regular spacing, the double staggered layout (\textit{U-C2} at a spanwise shift of $1.5D$) outperforms the staggered layout (e.g. \textit{U-C1} at a spanwise shift of $2.5D$). The layout series with a moderate uneven streamwise spacing, e.g. \textit{NU1-C1}, does not indicate any advantages, as it performs less well than the original layout series \textit{U-C1}. For the \textit{NU1-C2} series, very little influence of the spanwise shift is seen, so that it also does not provide any obvious advantages. The \textit{NU2-C1} series, at a zero shift, produces the lowest farm efficiency of all layouts. However, the power increases fast for a shift larger than $1D$, and is higher than any of the earlier discussed layouts (e.g.  \textit{U-C1},\textit{U-C2}, \textit{NU1-C1} and \textit{NU1-C2}), at a spanwise shift of $2.5D$. The highest farm efficiencies are measured for the layout series \textit{NU2-C2} and \textit{NU2-C3}. Interestingly, the maximum power of these layouts is not observed at the maximum spanwise shift of $1.5D$, which would result in more uniform spanwise distribution (the spanwise distribution of porous disk models would be uniform for a spanwise shift of $1.66D$). Instead, the maximum efficiency is reached at a smaller spanwise shift of $1D$, because of smaller wake losses, and possibly indicating that local flow accelerations due to the close spacing may play a role in this maximum performance. The layout series \textit{NU2-C3} and \textit{NU2-C3} with a spanwise shift of $1D$ also results in low turbulence intensity levels, reaching a value of $13-14\%$ at the end of the wind farm, such that these layouts are found to give the highest power output with a low level of unsteady loading. 

As seen in figures \ref{f:WF_NU1} and \ref{f:WFNU2_P}, for a zero spanwise shift, the uneven layouts show an alternating pattern of very low and high surrogate power values, due to the strong wake losses. However, the layouts with the highest power of each series show a relatively smooth asymptotic behavior of the surrogate power at the end of the wind farm, indicating that a fully-developed regime is being approached. To investigate the influence of layout on the value of the asymptote, figure \ref{f:WFAS} (b) presents the average power of row 16 to 19.

The \textit{U-C1} and \textit{U-C2} series show very little differences for the mean power at the end of the farm as a function of layout. This observation shows that all the improvements in power are made in the entrance region of the farm. These observations are in good agreement with the top-down models \cite{fra92,fra06,calaf2010large,meneveau2012top,stevens2017flow}, which assumes that the wind turbine forces are uniformly applied on the flow, and predict a power asymptote which is only dependent on the wind turbine density. However, the uneven streamwise layout series \textit{NU1} and \textit{NU2} show a significant variation of the power at the end of the farm, with the lowest value when the spanwise shift is zero. Strong wake losses can thus influence the entire farm and reduce the asymptote for non-uniform layouts. It is important to note that the power asymptote in the fully-developed regime can also decrease if the spanwise spacing would be increased (and consequently the streamwise spacing proportionally decreased), as the transverse wake expansion is limited and the area occupied by the wind farm becomes less optimally used \cite{yang2012computational}. Such an effect is taken into account in the coupled wake boundary layer (CWBL) model \cite{stevens2016generalized} using an effective coverage area that may be smaller than the actual area for wide spanwise spacings. However, this effect is not playing a role in the current experiments as the spanwise spacing is kept constant at a value of $S_y/D = 5$.

The maximum power for the layouts with a moderate uneven streamwise spacing \textit{NU1}, found for a spanwise shift of $2.5D$, reaches approximately the same value as for the \mbox{ \textit{U-C1}} and \mbox{\textit{U-C2}} series. Interestingly, the \textit{NU2} layout series can reach a slightly higher maximum value, with the highest power found for the \textit{NU2-C2} series and a spanwise shift of $1D$. It is important to consider that the measurement uncertainty of the strain gages is not negligible. However, at a spanwise shift of $\Delta_y /D=1$ the difference in power between layout \textit{NU2-C2} and \textit{U-C1} is larger than the estimated measurement uncertainty, and thus considered significant. These results indicate that the extreme non-uniform streamwise spacing of the \textit{NU2} layout series can have benefits for both the entrance region of the wind farm and the fully-developed regime.

\section{Conclusions}

An experimental parametric study of farm layout was performed with the micro wind farm model in the Corrsin Wind Tunnel. The instantaneous forces of all sixty porous disk models in the central three columns of the wind farm were measured for 56 different layouts. The mean surrogate power of each model and the estimated local turbulence intensity was used to find the most optimal layout. By keeping the area occupied by the wind farm constant for each layout, we are especially interested in finding the configuration with the highest farm efficiency, as defined by the ratio of power over occupied area. Furthermore, the temporal data acquisition capabilities of the porous disk models are used to assess the unsteady loading caused by turbulent scales significantly larger than the disk.

Three main layout series were considered, a series with a uniform streamwise spacing ($S_x/D = 7$), with a moderate alternating streamwise spacing ($S_x/D = 3.5$ and $S_x/D = 10.5$), and with an extreme alternating streamwise spacing ($S_x/D = 1.5$ and $S_x/D = 12.5$). For each series, layout variations are created by sliding specific rows in the spanwise direction.

The experiments resulted in a vast data-set of surrogate mean row power and local turbulence intensity for each layout, in controlled and documented conditions. For each series, the layout with the highest overall power, shows a relatively smooth decrease of the row power towards an equilibrium value at the end of the farm. 
This trend is in agreement with results in the literature \cite{stevens2014large,stevens2016effects}. The largest improvements in farm efficiency are created by the increase of surrogate power in the first half of the wind farm. All layouts with a uniform streamwise spacing approach approximately the same value at the end of the farm, in agreement with the top-down model \cite{fra92,calaf2010large}, which predicts a single power asymptote for a certain wind turbine density.

However, for the layouts with an alternating streamwise spacing, the mean power at the end of the farm shows a strong dependence on the spanwise shift. The lowest values are generally reached when the spanwise shift is zero, due to strong wake effects. For a moderate uneven streamwise spacing, the maximum power at the end of the farm is reached with a spanwise shift of $2.5D$, and is approximately the same as for a uniform layout. Interestingly, for an extreme uneven streamwise spacing, a slightly higher value is reached at the end of the farm (up to $\approx 5-6\%$) for a spanwise shift of $1D$. The layouts with an extreme uneven spacing were also found to measure the highest farm-aggregate surrogate power, which indicates advantages for both the entrance and the fully-developed region. These results indicate the possible beneficial role of local flow accelerations, similar to the results by McTavish et al. \cite{mctavish2014experimental} for three wind turbines. Such flow dynamics are not naturally included in analytical wake models. It would therefore be interesting to verify if analytical and numerical models predict similar trends as observed in these experiments. The experimental results can therefore be useful for future testing of wind farm models.

For each series, the layout with highest overall power, also results in the lowest unsteady loading. All of these layouts indicate a similar, slow progression of the unsteady loading, which levels off after approximately $11-13$ rows, and reaches a value of $TI \approx13-14\%$. For the less optimal layouts, the unsteady loading increases due to wake effects. 

Overall it is concluded that the layouts with an extreme alternating streamwise spacing can result in the highest surrogate power and a low unsteady loading if the spanwise shift is larger or equal to $1D$. Specifically the layout \textit{NU2-C2} with a spanwise shift of $1D$ showed the most optimal results. The disadvantage of the layouts with an extreme non-uniform spacing is that for certain wind directions the wake losses can become very large, as indicated in figure \ref{f:WFAS} for a zero spanwise shift. Future studies should explore in more detail the flow interactions and resulting beneficial effects of closely spacing small groups of wind turbines for a range of wind directions.

\section*{Acknowledgements}
Work is supported by ERC (grant no. 306471, the ActiveWindFarms project) and by NSF (grant OISE-1243482, the WINDINSPIRE project).

\bibliographystyle{unsrtnat}
\bibliography{allpapers}

\end{document}